\documentclass[journal=jctcce,manuscript=article,
layout=twocolumn]{achemso}
\usepackage[T1]{fontenc} % Use modern font encodings
\usepackage{graphicx}
\usepackage{multirow}
\usepackage{longtable}
\usepackage{dcolumn}% Align table columns on decimal point
\usepackage{bm}
\usepackage{amsmath}
\usepackage{amssymb}
\usepackage{txfonts}
\usepackage{booktabs}
\usepackage[version=3]{mhchem}
\usepackage{url}
\usepackage{here}
\usepackage{hyperref}
\usepackage{physics}
\usepackage[usenames,dvipsnames]{color}
\definecolor{Gray}{gray}{0.0}
\definecolor{lightGray}{gray}{0.35}
\newcommand{\VCo}{V$_\mathrm{Co}$}
\newcommand{\SG}{$P\bar{6}2m$}
\newcommand{\WO}{W$_\mathrm{O}$}
\newcommand{\rCo}{{\rm{Co}{}}}
\newcommand{\rC}{{\rm{C}{}}}
\newcommand{\rO}{{\rm{O}{}}}
\newcommand{\rH}{{\rm{H}{}}}
\newcommand{\re}{{\rm{e}{}}}

\author{Kenji Oqmhula}
\email{mwkokk1907@icloud.com}
\affiliation{
  School of Information Science, Japan Advanced Institute of 
  Science and Technology, Asahidai 1-1, Nomi, Ishikawa, 923-1292, Japan
}
\author{Takahiro Toma}
\affiliation{
  School of Information Science, Japan Advanced Institute of 
  Science and Technology, Asahidai 1-1, Nomi, Ishikawa, 923-1292, Japan
}
\author{Ryo Maezono}
\affiliation{
  School of Information Science, Japan Advanced Institute of 
  Science and Technology, Asahidai 1-1, Nomi, Ishikawa, 923-1292, Japan
}

\author{Kenta Hongo}
\email{kenta_hongo@mac.com}
\affiliation{
  Research Center for Advanced Computing Infrastructure, JAIST,
  1-1 Asahidai, Nomi, Ishikawa 923-1292, Japan
}
\date{\today}
\title{
  First-Principles-Based Insight into Electrochemical Reactivity in a Cobalt-Carbonate-Hydrate Pseudocapacitor
}
\begin{document}
%%%%%%%%%%%%%%%%%%%%%%%%%%%%%%%%%%%%%
%\begin{tocentry}
%  \centering
%  \includegraphics[keepaspectratio, width=8cm]{TOC.pdf}
%  \label{For Table of Contents Only}
%\end{tocentry}
%%%%%%%%%%%%%%%%%%%%%%%%%%%%%%%%%%%%%
\begin{abstract}
Cobalt carbonate hydroxide (CCH) is a pseudocapacitive material
with remarkably high capacitance and cycle stability.
Previously, it was reported that CCH pseudocapacitive materials are orthorhombic in nature.
Recent structural characterization has revealed that they are hexagonal in nature;
however, their H positions still remain unclear.
In this work, we carried out first-principles simulations to identify the H positions.
Through the simulations,
we could consider various fundamental deprotonation reactions inside the crystal
and computationally evaluate the electromotive forces (EMF) of the deprotonation ($V_\mathrm{dp}$).
Compared with the experimental potential window of the reaction
($< 0.6$ V (vs. saturated calomel electrode (SCE))),
the computed $V_\mathrm{dp}$ (vs. SCE) value ($3.05$ V) was beyond the potential window,
indicating that deprotonation never occurred inside the crystal.
This may be attributed to the strongly formed hydrogen-bonds (H-bonds) in the crystal,
thereby leading to the structural stabilization.
We further investigated the crystal anisotropy in an actual capacitive material
by considering the growth mechanism of the CCH crystal.
By associating our X-ray diffraction (XRD) peak simulations with experimental structural analysis,
we found that the H-bonds formed between CCH
$\{(\bar{1}\bar{1}\bar{1}), (2\bar{1}\bar{1}), (2\bar{1}1)\}$ planes
(approximately parallel to $ab$-plane) can result in 1-D growth (stacked along with $c$-axis).
This anisotropic growth controls the balance
between the total ``non-reactive'' CCH phases (inside the material)
and ``reactive'' hydroxide (Co(OH)$_2$) phases (surface layers);
the former stabilizes the structure,
whereas the latter contributes to the electrochemical reaction.
The balanced phases in the actual material can realize high capacity and cycle stability.
The results obtained highlight the possibility of regulating the ratio of the CCH phase
versus the Co(OH)$_2$ phase by controlling the reaction surface area.
\end{abstract}
\maketitle
%SSSSSSSSSSSSSSSSSS
\section{Introduction}
\label{sec.morphology}
%SSSSSSSSSSSSSSSSSS
For clean energy applications, pseudocapacitors with Co-based anodes can be applied to
large-scale power storage devices with excellent characteristics, such as fast charge--discharge,
long life, and high capacity.~\cite{2020FLE, 2021LIA}
Because of the balance between high capacity and cycle stability,
cobalt carbonate hydroxide (CCH)~\cite{2017LIN,2021SHU}
has recently attracted increasing attention
compared to conventional Co-based anode materials,
such as cobalt oxide (Co$_3$O$_4$),~\cite{2009DEN,2020MIR}
cobalt hydroxide (Co(OH)$_2$)~\cite{2010KON,2017DEN},
mixture of cobalt carbonate and cobalt oxide (CoCO$_3$/CoO)~\cite{2015JI},
and hybrid materials of ionic liquid and Co(OH)$_2$.~\cite{2013CHO}
In fact, it has been reported that a pristine CCH anode exhibits low capacity.~\cite{2015GHO}
In addition, the ``high-capacity CCH anode'' is not obtained from CCH itself,
but from Co(OH)$_2$ formed in the CCH by calcination,~\cite{2015GHO,2017LIN,2021LIU}
although the CCH contributes to high cycle stability
due to its robust/resilient structure against the electrochemical reaction.
XRD experiments have suggested the existence of mixed phases
consisting of CCH and Co(OH)$_2$.~\cite{2015GHO, 2017LIN}
In charge--discharge cycles of ``high-capacity CCH anode,''
Co(II) and Co(III) states have been observed alternatively,
suggesting that the Co(OH)$_2$ is deprotonated to CoOOH,
followed by deprotonation to CoO$_2$, and {\it vice versa}.
The electrochemical reaction in Co(OH)$_2$ has been investigated
by first-principles simulations~\cite{2009DEN},
thus providing atomistic insight into the detailed mechanism.
Analogous to this, it has been proposed that
the abovementioned two step reaction is attributed to
the high capacity of the CCH anode.~\cite{2015GHO,2017LIN,2021LIU}
The abovementioned reaction mechanism can be considered as plausible,
but hypothetical, {\it i.e.,}
there is no evidence of electrochemical reactions
in the CCH anode at the atomic level
because the CCH crystal structure
still remains unclear~\cite{2019BHO,2021SHU};
therefore, similar first-principles/atomistic level simulations have not been applied yet.

\vspace{2mm}
In the high-capacity CCH anodes, their morphologies strongly affect their capacitance,
{\it i.e.,} capacity and cycle life.
The best capacitance, so far, has been achieved for the CCH anode
with an umbrella-like morphology controlled by the crystal growth time.~\cite{2017LIN, 2021SHU}
This anode material has a well-controlled 1-D anisotropic morphology,
including well-balanced CCH and Co(OH)$_2$ phases;
the former is responsible for the structural stability
({\it i.e.,} long life cycle),
while the latter is responsible for high capacity.
Therefore, morphology of CCH anode balancing CCH and Co(OH)$_2$ phases
is critical for achieving outstanding capacitance properties.
The morphology of CCH can be controlled by the crystal growth direction;
the more 1-D anisotropic morphology results in less CCH phase/more Co(OH)$_2$ phase,
thereby leading to the larger reactive area, and hence, higher capacity.
This anisotropy can be controlled by the packing ratio of carbonate ions in the crystal.
However, the reason behind carbonate ion's contribution
in crystal growth is still unclear, because there was no detailed
information on the atomic positions in the CCH crystal structure.

\vspace{2mm}
Thus, accurately identifying the CCH crystal structure is the first step
toward the deeper understanding of ``high-capacity'' CCH anodes,
which will be helpful for designing higher-capacity CCH anodes.
For instance, if the position of carbonate ions is known,
we should be able to consider
the deprotonation site of the hydroxide phase formed by calcination.
The CCH compound includes carbonate and hydroxide ions as well as hydrated water;
its crystal structure is much more complex than those of oxides and hydroxides.
Previously, an XRD analysis assuming an orthorhombic lattice identified CCH compound
as Co(OH)(CO$_3$)$_{0.5}$ $\cdot$ 0.11H$_2$O(JCPDS 48-0083).~\cite{1992POR}
However, recent sophisticated experiments suggested that CCH is a Co$_6$(CO$_3$)$_2$(OH)$_8$H$_2$O
with a hexagonal lattice.~\cite{2019BHO,2021SHU}
The crystal system has been identified,
but the detailed information about atomic positions
is yet to be identified:
it is quite difficult to identify hydrogen
(by neutron diffraction) and Co vacancy positions.

\vspace{2mm}
This study aims
(i) to accurately identify the CCH crystal structure and
(ii) to elucidate the relationship between the material morphology and the crystal growth;
the first addresses the ``bulk'' aspect,
whereas the second addresses the ``material'' aspect in a pseudocapacitor.
Based on the resulting structures,
we systematically investigated the capacitance properties of the CCH anodes (reductant),
referring to those reported in previous studies.
The EMF of the CCH anodes was evaluated by first-principles simulations
to investigate the electrochemical reactivity.
We found that the CCH crystal forms H-bond networks
that can contribute to the cyclic stability;
the H-bond inhibits the deprotonation reaction in the CCH crystal.
Based on this finding, we considered the proton conduction mechanism
and capacitance in the CCH anode.
Eventually, we may conclude that the reaction site is not attributed to the CCH phase,
but to the Co(OH)$_2$ phase; CCH can be considered as the precursor of the Co(OH)$_2$ phase,
with wire-like morphology.
To identify the crystal planes that contribute toward the crystal growth,
we performed XRD peak simulations on the resultant crystal structure
with the hexagonal lattice and assigned the plane indices.
We found that the H-bonds are formed between the
$\{(\bar{1}\bar{1}\bar{1}), (2\bar{1}\bar{1}), (2\bar{1}1)\}$
planes that contribute toward the 1-D crystal growth.

\vspace{2mm}
This paper is organized as follows:
Section~\ref{method} starts with modeling of the CCH crystal structure
to identify hydrogen and Co defect positions therein;
to evaluate the EMF of the reaction using first-principles simulations,
we hypothesize the CCH electrode reaction (deprotonation)
based on the structural modeling of the oxide (dehydrogenation location identification),
followed by the present first-principles simulations.
In Section~\ref{result}, the results of this work are presented, i.e.,
identification of the reductant structure,
oxidant structure to evaluate the EMFs of the deprotonation,
and the XRD simulation to assign the peaks to correct Miller indices.
According to the results obtained,
section~\ref{discussion} provides new insights
into the storage performance of CCH-based electrodes.
The conclusion drawn from the findings are provided in Section~\ref{conclusion}.

%%%%%%%%%%%%%%%%
\section{Methodologies}
\label{method}
%SSSSSSSSSSSSSSSSSS
\subsection{Outline}
%SSSSSSSSSSSSSSSSSS
Exhaustive first-principles simulations for all possible atomic configurations were carried out,
and CCH structures in reduced and oxidized forms were identified as the most stable structures.
The electrochemical reactivity of the CCH anode was evaluated from the EMF value,
which is defined as the energy difference between the reductant and oxidant.
Furthermore, their structural changes provide critical information
for understanding the cycle performance
because they correlate well with each other.~\cite{2017DEN, 2008SIM, 2018KIM}
The structural and computational modeling are outlined as follows:
%------------------
\begin{itemize}
\setlength{\parskip}{-4pt}
%------------------
\item[(i)] According to the recent experiments on structural
  characterization of CCH anodes,~\cite{2019BHO,2021SHU}
  we considered the CCH reductant to be Co$_6$(CO$_3$)$_2$(OH)$_8$H$_2$O
  in the hexagonal \SG{} space group (No. 189);
  however, its Co defect ({\VCo}) and
  H positions have not been determined experimentally.
  (See Subsection~\ref{reductant}.)
  This indicates that there are several possible
  arrangements of {\VCo} and H positions therein.
  First-principles simulations are well-known
  for identifying unknown atomic positions
  in disordered structures.~\cite{2020TOM}
  Therefore, we exhaustively applied first-principles simulations
  to all the possible arrangements,
  identifying the reductant structure as
  the most stable one.
  (See Subsection~\ref{simulations}.)
\item[(ii)]
  To theoretically elucidate the electrochemical properties of the CCH anode,
  the most reliable approach is first-principles reaction dynamics simulations.
  However, due to its high computational cost,
  such simulations are not feasible.
  Instead, the present study adopted a static approach, i.e.,
  we evaluated the EMF.
  To evaluate the oxidant energy, the oxidant structure should be modeled,
  but the modeling involves assuming a plausible electrochemical reaction.
  Before modeling the oxidant structure,
  we review the electrochemical reactions, which are
  hypothetically assumed as deprotonation reaction.~\cite{2015GHO, 2017LIN, 2021SHU, 2021LIU}
  (See Subsection~\ref{reaction}.)
\item[(iii)]
  According to the hypothetical reaction mechanism,
  we can model the oxidant structure by deprotonating
  the reductant structure.
  (See Subsection~\ref{motiveforce}.)
  By applying the first-principles simulations
  to both the reductant and oxidant structures,
  the EMF ($V_\mathrm{dp}$)
  can be calculated from their energy difference computed
  using the first-principles geometry optimization.
  (See Subsection~\ref{simulations}.)
  Upon comparing the predicted $V_\mathrm{dp}$
  with the experimental values, we verified the reaction hypothesis.
\item[(iv)]
  Actual CCH anodes have 1-D wire structures rather than
  the 3-D bulk structure.~\cite{2015GHO, 2017LIN, 2021SHU, 2021LIU}
  Therefore, we considered the anisotropic morphologies of the CCH anodes
  to understand ``actual'' capacitor properties
  from a more realistic viewpoint.
  The crystal growth mechanism can govern the morphology.
  The anisotropy of the crystal growth can be
  measured by the XRD pattern;
  the crystal grows toward crystal planes that
  corresponds to higher peak intensities
  in its XRD pattern.~\cite{2003XU}
  In fact, previous characterizations of the CCH anodes
  were incorrect because they were characterized by
  assuming a lattice structure
  with an incorrect space group.~\cite{1992POR}
  Therefore, we perform XRD simulations~\cite{2011MOM}
  for our predicted CCH reductant structure
  to obtain the accurate attribution of the XRD patterns to crystal planes.
  (See Subsection~\ref{xrd}.)
  This revised XRD attribution can lead to a correct crystal growth mechanism.
  In addition, the detailed structural information about carbonate ions and
  surrounding atoms clarify the interactions between them.
  This detailed information can provide new insights
  into the morphology of the CCH anode and
  hence its electrochemical reactivity
  governing the capacitor properties, i.e.,
  high capacity and cycle stability.
\end{itemize}

%SSSSSSSSSSSSSSSSSS
\subsection{Modeling of CCH reductant structure}
\label{reductant}
\begin{figure}[]
  \centering
  \includegraphics[width=0.8\hsize]{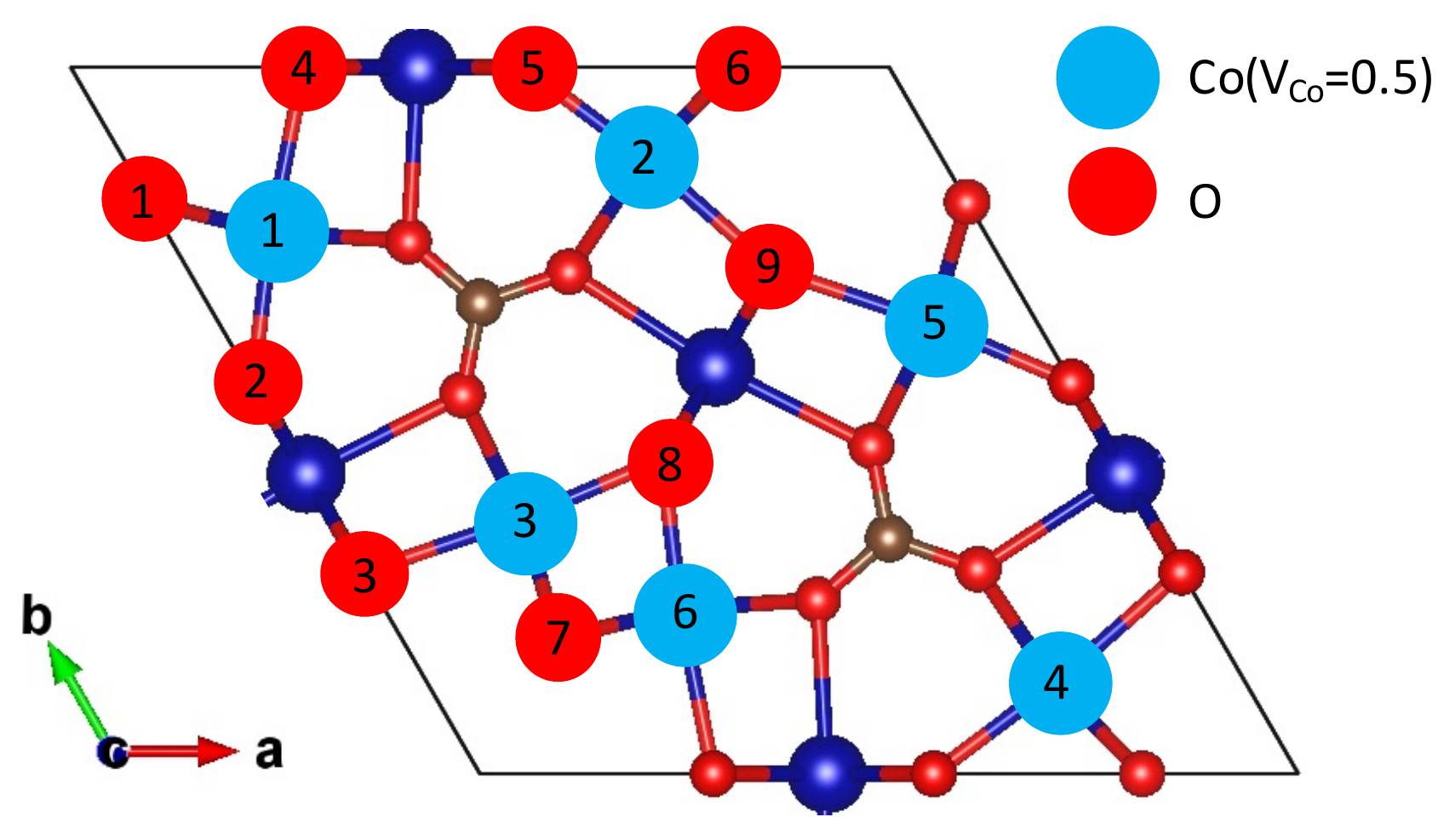}
  \caption{
  \label{fig.unitcell}
  Unit cell in the $ab$ plane. Atomic positions colored by
  light blue and red are 
  six Co-defective (\VCo{}) sites with site occupancy number of 0.5
  and nine O sites that bond with H atoms, respectively.
  Three of the \VCo{} sites
  and one of the O sites exist as vacancies and structural water,
  respectively.
  }
\end{figure}

%SSSSSSSSSSSSSSSSSS
Figure~\ref{fig.unitcell} shows a unit cell of Co$_6$(CO$_3$)$_2$(OH)$_8$H$_2$O
obtained from experimental characterization~\cite{2019BHO},
which is the starting point of our structural modeling.
Its lattice information (lattice parameters and space group) is identified,
but atomic positions therein are incomplete from the viewpoint of 
the Co-defect (\VCo) and H positions.
The atomic positions of Co, O, and C are identified,
but there exist six Co positions with site occupancy numbers of 0.5 (light blue).
This means that three Co positions are fully occupied and
the remaining three are \VCo positions.
Ten H atomic positions are categorized into eight OH sites and
one structural water (H$_2$O) site.
Through the first-principles simulations shown later,
the structural model of reductant was determined by
choosing the most stable structure from all the possible arrangements
of the \VCo{} and H positions.
As shown below, we identify (i) the \VCo{} and (ii) H positions
considering physicochemically reasonable arrangements.
%--------------------------------
%--------------------------------

\vspace{2mm}
(i) Identification of the \VCo{} positions:
The vacancy patterns can be divided into two categories:
(1) delocalized and (2) localized ones.
According to the symmetry of the \VCo{} patterns,
we have two delocalized patterns,
(1-1) V$_{\mathrm{Co}}$(2/3/5) and (1-2) V$_{\mathrm{Co}}$(1/2/3),
and two localized ones,
(2-1) V$_{\mathrm{Co}}$(2/3/4) and (2-2) V$_{\mathrm{Co}}$(3/5/6),
where the three \VCo{} positions are described in parentheses
as numbers in Figure~\ref{fig.unitcell}.

\begin{figure*}[hbtp]
  \centering
  \includegraphics[width=0.8\hsize]{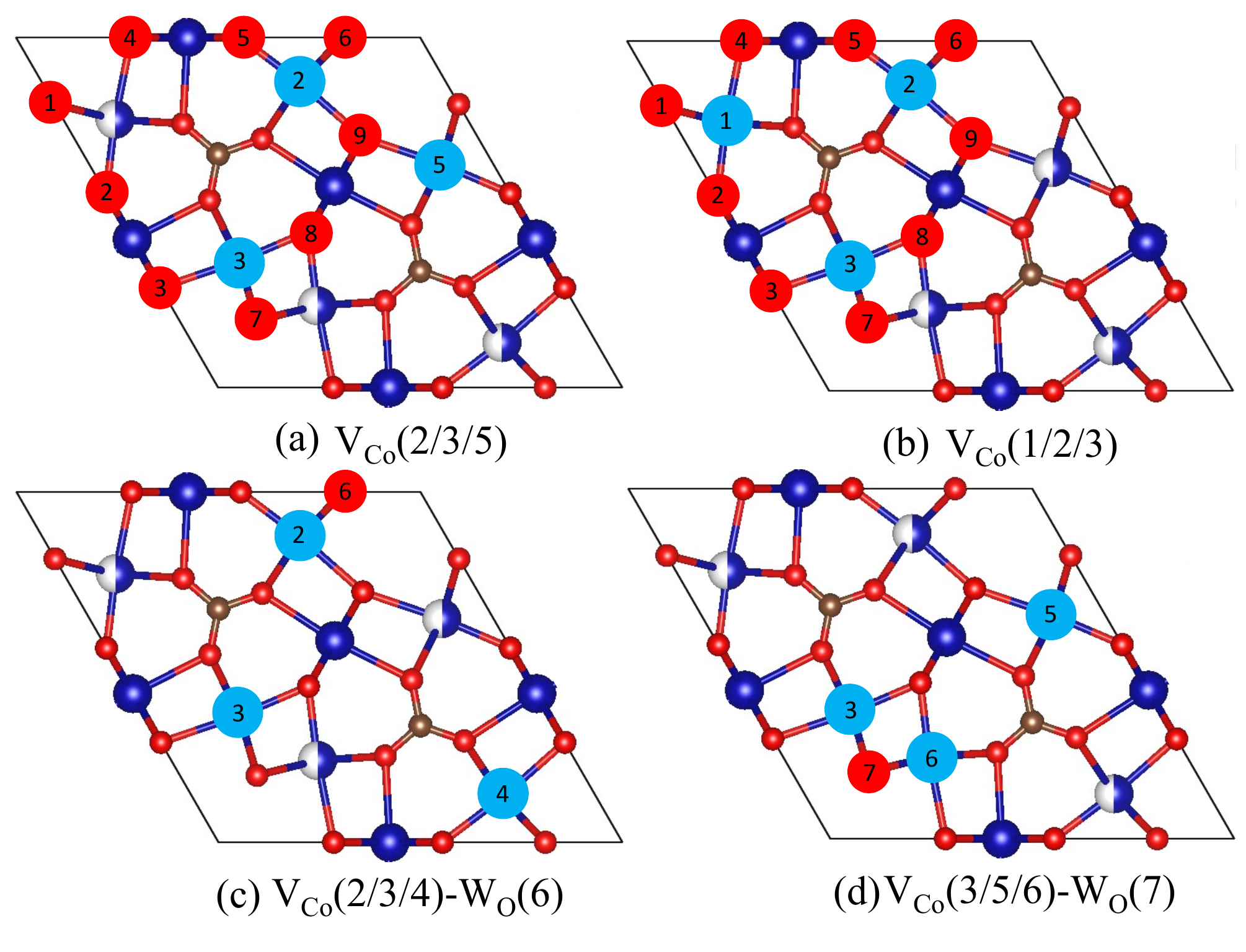}
  \caption{
  \label{fig.structure}
  Possible structural models of \VCo($i/j/k$)-\WO($l$),
  where \VCo{} and structural water positions are
  colored by light blue and red, respectively:
  (a) nine possible patterns, \VCo(2/3/5)-\WO($l$)~$(l=1\sim 9)$,
  (b) three possible patterns, \VCo(1/2/3)-\WO($l$)~$(l=1,2,3)$,
  (c) one unique pattern, \VCo(2/3/6)-\WO(6),
  and (d) \VCo(3/5/6)-\WO(7).
  }\end{figure*}

\vspace{2mm}
(ii) Identification of the H positions:
For each of the above four \VCo{} patterns (1-1)$\sim$(2-2),
the H positions are automatically determined
if one H$_2$O position is chosen from nine O positions,
\WO{}$(l)~(l=1\sim9)$, appearing in Figure~\ref{fig.unitcell}.
Furthermore, as explained below,
we consider their symmetries
and impose constraint conditions from the viewpoint of
the local charge-neutrality that isolated anions
are ruled out in the reductant,
and thus, 14 candidate structures are left,
described as \VCo($i/j/k$)-\WO($l$):
\begin{itemize}
\item[(1-1)] \VCo(2/3/5)-\WO{}$(l)~(l=1\sim9)$:
  As can be seen from Figure~\ref{fig.structure} (a),
  this case allows all the nine O sites
  because neither symmetry nor local charge-neutrality rule out
  any \WO{} possibility.
\item[(1-2)] V$_{\mathrm{Co}}$(1/2/3)-\WO$(l)~(l = 1,2,3)$:
  Figure~\ref{fig.structure} (b) shows \VCo{}(1/2/3)
  has three-fold symmetry; therefore,
  \WO{}$(l)$ are classified into three
  equivalences:
  \{\WO(1), \WO(6), \WO(7)\},
  \{\WO(2), \WO(5), \WO(8)\},
  and \{\WO(3), \WO(4), \WO(9)\}.
  Therefore, we may consider only three inequivalent \WO$(l)~(l=1,2,3)$.
\item[(2-1)] \VCo(2/3/4)-\WO(6):
  Looking at Figure~\ref{fig.structure} (c) and then
  comparing every O valency with the \VCo valencies,
  the local charge-neutrality restricts
  possible candidates for the structural water site
  to only one O site, \WO(6).
  Otherwise, O6 atom becomes isolated OH$^-$.
\item[(2-2)] \VCo(3/5/6)-\WO(7):
  Similar to the \VCo(2/3/4) case, the \WO(7)
  is only the possible O site for the structural water,
  looking at the valencies of Co3 and Co6 vacancies.
  (See Figure~\ref{fig.structure} (d).)
\end{itemize}
%--------------------------------
%--------------------------------

\vspace{2mm}
Note that when optimizing these initial structural models
by first-principles simulations (Subsection~\ref{simulations}),
the degree of spatial freedom is restricted such that
H atoms in OH$^-$/H$_2$O are placed parallel/perpendicular to
the $ab$ plane, which avoids undesirable steric hindrance.

%%%%%%%%%%%%%%%%%%%%%%%%%%%%%%%%%%
\subsection{Hypothesis of deprotonation\\reaction and EMF}
\label{reaction}
%%%%%%%%%%%%%%%%%%%%%%%%%%%%%%%%%%
\vspace{2mm}
The electrochemical reaction of the CCH anode has been
inferred from experiments based on
cyclic voltammetry (CV) measurements.~\cite{2017LIN}
The CV measurements have provided
the reaction EMF by observing
the CV peaks appearing in the potential window,
$V < 0.6$V versus the SCE potential reference.
The result suggests the following
deprotonation reaction:
%--------------------------------
\begin{equation}
  \label{eq:CCH}
  \mathrm{Co(OH)}_2 \rightleftharpoons \mathrm{CoOOH +H^+ + e^-}.
\end{equation}
%--------------------------------
Equation~\eqref{eq:CCH} is
an electrochemical (elementary) reaction of Co(OH)$_2$ rather than CCH itself,
indicating that Co(OH)$_2$ forms in the CCH anode.
The corresponding half reaction in the CCH anode
for $n$ deprotons per unit cell ($1 \le n \le 6$)
is given as:
%--------------------------------
\begin{align}
  \label{eq:anode}
  &\rCo_6\rC_2\rO_{15}\rH_{10} + n\rO\rH^- =\\
  &\rCo_6\rC_2\rO_{15}\rH_{10-n} + n\rH_2\rO + \textit{n}\re^- 
  + V_\mathrm{CCH}~\rm(vs. SHE), \notag
\end{align}
%------------------
where $V_\mathrm{CCH}\rm(vs. SHE)$ is the CCH electrode potential
versus the standard hydrogen electrode (SHE).
Hereafter, all the electrochemical potentials are measured referring to SHE.
For $n = 6$, all the Co$^{+2}$ ions are oxidized to Co$^{+3}$,
with the reaction rate being 1.
The cathodic half-reaction is assumed to be the Pt electrode reaction
in an alkaline solution (KOH):
%--------------------------------
\begin{equation}
  \label{eq:cathode}
  n\rH_2\rO + n\re^- = n\rO\rH^- + \frac{n}{2}\rH_2
  + V_\mathrm{Pt}~\rm(vs. SHE),
\end{equation}
%--------------------------------
where $V_\mathrm{Pt}$ is the Pt electrode potential.
Eventually, the overall reaction is obtained:
%--------------------------------
\begin{equation}
  \label{eq:all}
  \rCo_6\rC_2\rO_{15}\rH_{10} =
  \rCo_6\rC_2\rO_{15}\rH_{10-n} + \frac{n}{2}\rH_2
  + V_\mathrm{dp}~\rm(vs. SHE),
\end{equation}
%--------------------------------
where $V_\mathrm{dp}$ is the deprotonation potential.

\vspace{2mm}
Herein, we establish relationships between the electrochemical potentials.
The deprotonation potential in Eq.~\eqref{eq:all} is a sum of
the anodic potential in Eq.~\eqref{eq:anode} and the cathodic one
in Eq.~\eqref{eq:cathode}:
%------------------
\begin{equation}
  \label{eq:CCH_vol}
  V_{\mathrm{dp}} = V_{\mathrm{CCH}} + V_{\rm{Pt}}~\mathrm{(vs. SHE)}
\end{equation}
%------------------
The $V_\mathrm{CCH}$ has been measured by the CV experiment
as the peak of EMF referring to the SCE potential.~\cite{2007LID}
We convert the experimentally set $V_\mathrm{CCH}$ (vs. SCE)
to a computationally tractable $V_\mathrm{CCH}$ (vs. SHE) via
%------------------
\begin{equation}
  \label{eq:SCE}
  V_{\mathrm{CCH}}~\mathrm{(vs.SHE)} = V_{\mathrm{CCH}}~\mathrm{(vs.SCE)}
  + V_{\mathrm{SCE}}~\mathrm{(vs.SHE)},
\end{equation}
%------------------
where $V_\mathrm{SCE}~\mathrm{(vs.SHE)}$ is
the SCE electrode potential measured by SHE.
Substituting Eq.~\eqref{eq:CCH_vol} with experimental values
($V_\mathrm{SCE} = 0.2681$ V and $V_\mathrm{Pt} = -0.8277$ V)~\cite{2007LID}
into Eq.~\eqref{eq:SCE}, we obtain the EMF (vs. SCE) for the CCH anode
as
%------------------
\begin{equation}
  \label{eq:CCH_vs_SCE}
  V_\mathrm{CCH}~\mathrm{(vs.SCE)}
  = \left(V_{\mathrm{dp}} - 1.0958\right)\mathrm{V~(vs.SHE)}.
\end{equation}
%------------------
Thus, the $V_\mathrm{CCH}~\mathrm{(vs.SCE)}$ value can be evaluated
from the $V_\mathrm{dp}~\mathrm{(vs.SHE)}$ one that can be
computed by first-principles simulations, as shown in the next subsection.

%------------------
\subsection{First-principles evaluation of deprotonation EMF and oxidant modeling}
\label{motiveforce}
%------------------
The $V_\mathrm{dp}$ value can be converted into
the deprotonation energy, $\Delta E_\mathrm{CCH}$,
for $n$ deprotons, using
%------------------
\begin{equation}
  \label{eq:V}
  \Delta E_{\rm{dp}} = {nF}\cdot V_\mathrm{dp},
\end{equation}
%------------------
where $F$ is Faraday's constant.
Referring to the overall reaction in Eq.~\eqref{eq:all},
the $\Delta E_\mathrm{CCH}$ is given as
%------------------
\begin{equation}
  \label{eq:E}
  \Delta E_\mathrm{dp} = E_{\rm{CCH}} - \left[ E_{{\rm{CCH}}-n{\rm{H}}}
  + \frac{n}{2}E_{\rm{H}_2}\right],
\end{equation}
%------------------
where $E_\mathrm{CCH}$ and $E_{{\mathrm{CCH}}-n\mathrm{H}}$ are
the total energies of the reductant (CCH)
and the oxidant (CCH$-n$H) with $n$ deprotons, respectively,
and $\frac{\textit{n}}{2}E_{\rm{H}_2}$ is
the chemical potential of the hydrogen atom.

\begin{figure}[htb]
  \centering
  \includegraphics[width=0.95\hsize]{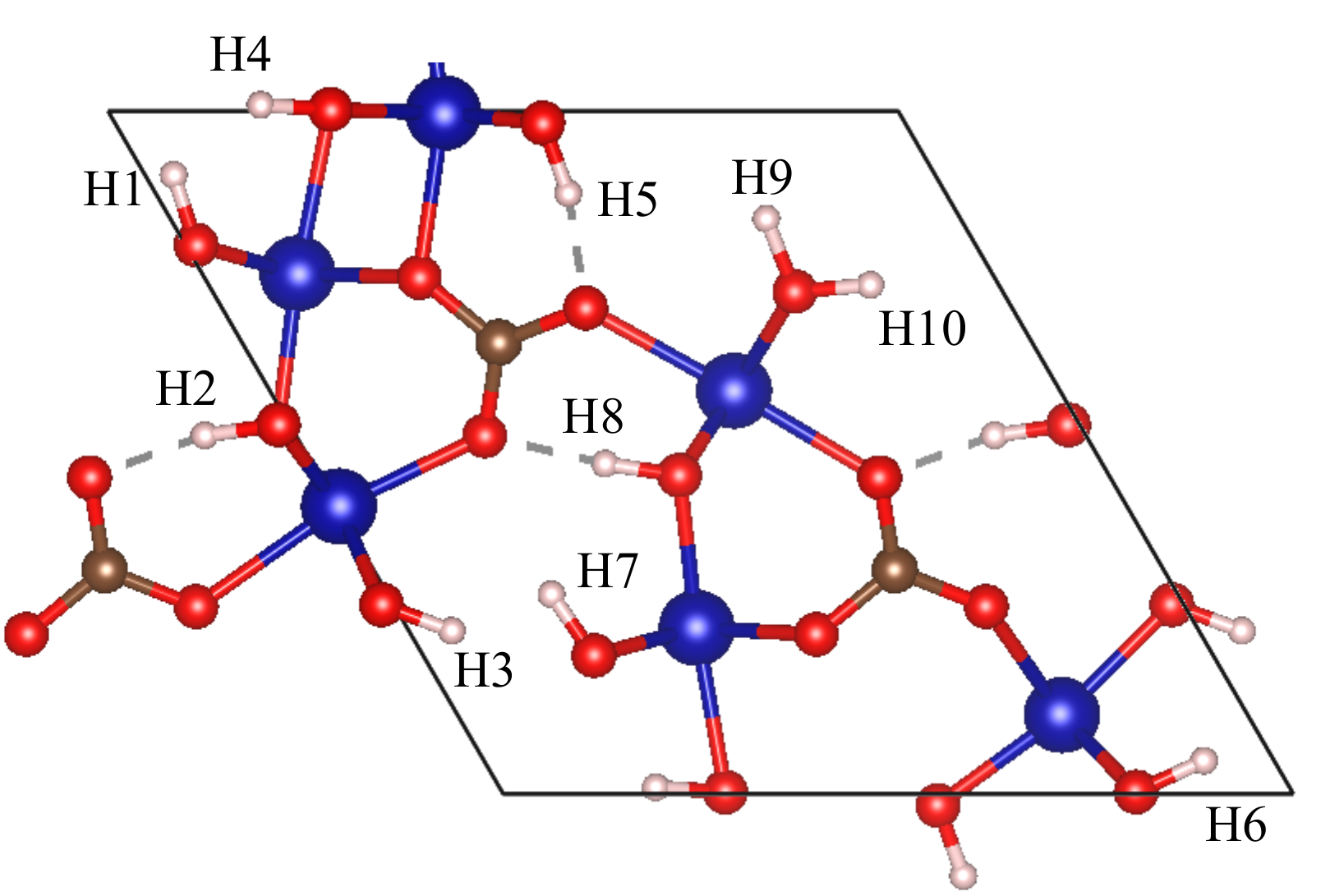}
  \caption{
  Deprotonation sites (H1$\sim$H10) in the unit cell on the $ab$ plane.
  Co, O, C, and H atoms are colored by blue, red, brown, and white,
  respectively.
  \label{fig.hyd_position}
  }
\end{figure}

\vspace{2mm}
Although the above three energies are computable, as shown in the next subsection,
the evaluation of $E_{{\rm{CCH}}-n{\rm{H}}}$ involves
modeling the oxidant structure.
In this study, we model the oxidant by
removing one hydrogen atom $(n=1)$ from the most stable reductant, \VCo{}(2/3/5);
ten possible deprotonation sites are shown in Figure~\ref{fig.hyd_position}.
For our purpose, as shown later, the $n=1$ case suffices to
evaluate $V_\mathrm{dp}$.
Note that our concern is whether or not the deprotonation reaction
occurs inside the CCH bulk.
In other words, we investigate
whether or not $V_\mathrm{CCH}$ lies in the range of
the experimental potential window, $V_\mathrm{exp} < 0.6$ V (vs. SCE).
%------------------
%------------------

%------------------
\subsection{First-principles geometry \\optimization}
\label{simulations}
%------------------
Our first-principles simulations based on
density functional theory (DFT)
were performed using
Vienna ab initio simulation package (VASP)~\cite{1996KRE,1999KRE}.
The valence electron-core interactions were described
by projector augmented wave (PAW),~\cite{1994BLO}
where the valence electrons were treated as Co(3s$^7$4s$^2$),
O(2s$^2$2p$^4$), C(2s$^2$2p$^2$), and H(1s$^1$).
We adopted the GGA-PBE exchange-correlation functional~\cite{1996PER}
and applied the GGA+U method to describe the localized Co-$3d$ orbitals
with the Hubbard correction value $U_{\rm{eff}}=6.1~\rm{eV}$
obtained from the value applied to cobalt oxide ~\cite{2007WDO};
we also added the Grimme D3 correction~\cite{2010GRI} with
the Beck-Johnson damping~\cite{2006JOH} to the GGA+U functional
to reproduce the noncovalent interactions.
The cutoff energy of the plane-wave basis expansion
was set to 700 eV;
a k-point mesh of $2\!\times 2\!\times 6$ was used
for integration in reciprocal lattice
space using the Monkhorst-Pack scheme~\cite{1976MON};
These values were found to converge within 1 meV/atom.
The convergence criteria for the geometry optimization
were set to $1\times 10^{-6}$ eV and 0.01 eV/{\AA}
for energy and force, respectively.
All the lattice parameters were fixed to be the experimental values~\cite{2019BHO}
during the geometry optimization,
i.e., only the atomic positions were optimized
by our first-principles simulations.
To accelerate the convergence in the geometry optimization,
we first optimized the H positions, followed by all the atomic positions,
which suppress a high degree of spatial freedom in the geometry optimization.

%------------------
\subsection{XRD pattern prediction}
\label{xrd}
%------------------
Previously, XRD peaks of the CCH electrodes were incorrectly
assigned to crystal plane indices because the CCH crystal structure
was assumed to be an orthorhombic lattice.{blue}~\cite{1992POR}
To re-assign the XRD peaks to plane indices of the hexagonal lattice,
we performed XRD simulations
for the computationally determined CCH hexagonal structure
using the VESTA software.~\cite{2011MOM}
Since the strongest XRD peaks indicate
the direction of the crystal growth,
new XRD alignments can provide a new insight into
1-D anisotropic CCH electrodes exhibiting
high capacity.~\cite{2017LIN, 2021LIU}

%%%%%%%%%%%%%%%%%%%%%%%%%%%%%%%%%%
\section{Results}
\label{result}
%sssssssssssssssssssssssssssssssssssssssssssssssssssss
\subsection{Identification of reductant}
\label{reductant}
\begin{figure*}[htb]
  \centering
  \includegraphics[width=0.8\hsize]{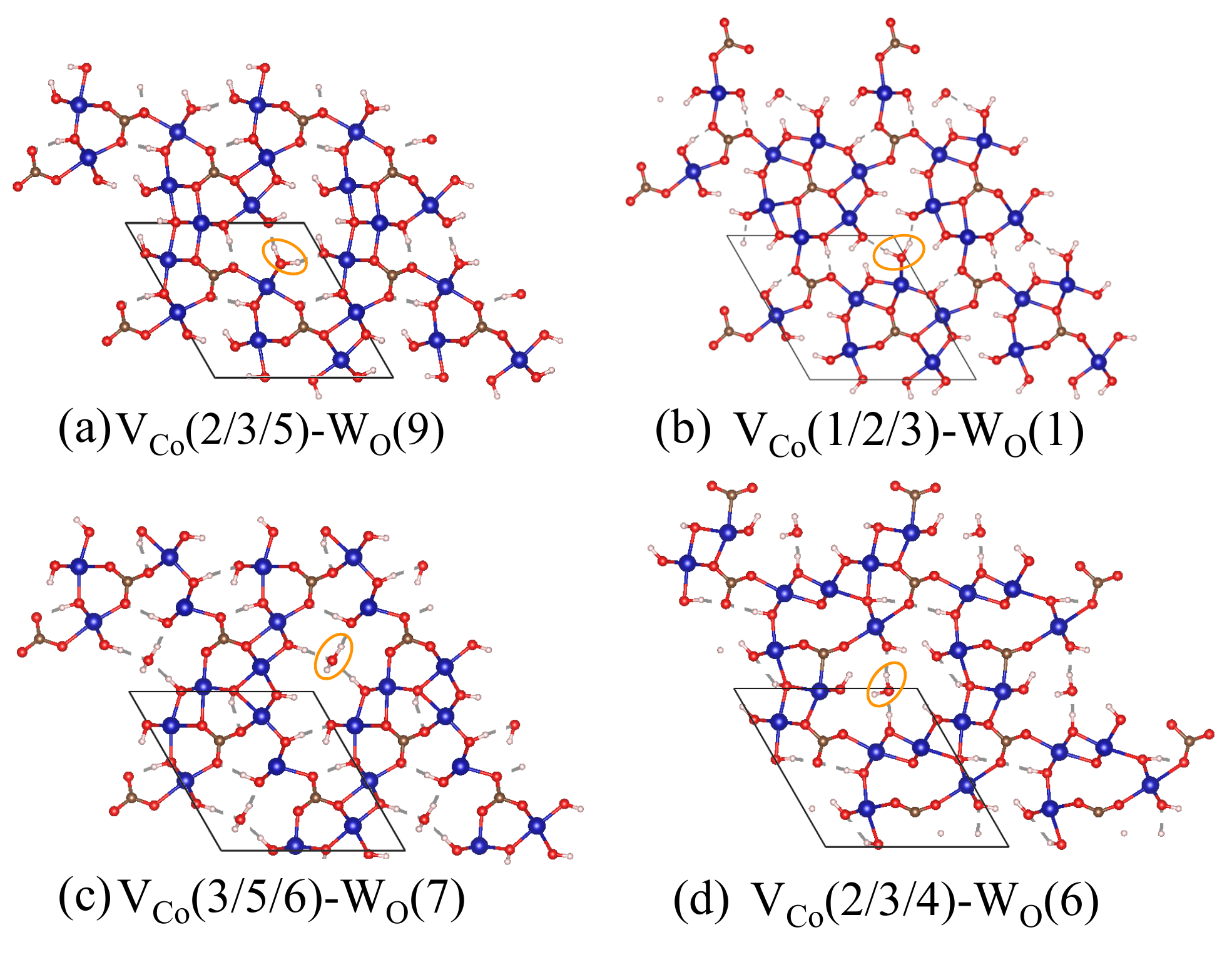}
  \caption{
    \label{fig.reductants}
    Four reductant structures with delocalized (1-1/1-2)
    and localized (2-1/2-2) arrangements of 
    Co vacancies, each of which is the most stable
    among their categories:
    (a) (1-1) \VCo(2/3/5)-\WO(9),
    (b) (1-2) \VCo(1/2/3)-\WO(1),
    (c) (2-1) \VCo(3/5/6)-\WO(7),
    and (2-2) \VCo(2/3/4)-\WO(6).
  Positions of structural waters are encircled
  by orange-colored ellipses.
  }
\end{figure*}

%sssssssssssssssssssssssssssssssssssssssssssssssssssss
To identify the reductant structure,
we performed the DFT simulations for all the possible arrangements
of \VCo{} and H$_2$O, i.e.,
14 patterns of \VCo($i/j/k$)-\WO($l$) shown in Figure~\ref{fig.structure},
followed by exploring the most stable one.
Table~\ref{tab.opt_hyd} lists four pairs of the \VCo{} and \WO{} positions,
and each of them provide the most stable structure among each category.
The most stable structure among the four arrangements is set to be zero energy;
the corresponding structures are shown in Figure~\ref{fig.reductants}.
The most stable structure was (1-1) \VCo(2/3/5)-\WO(9),
where the \VCo{} positions are delocalized, as shown in Figure~\ref{fig.reductants}(a).
This structure is stabilized by three hydrogen bonds:
two are formed between two H atoms in OH and two O atom in CO$_3^{2-}$,
and one is formed around \VCo{} surrounded by two CO$_3^{2-}$.
The second most stable structure is (1-2) \VCo(1/2/3)-\WO(1),
with only one hydrogen bond formed around one CO$_3^{2-}$.
which is shown in Figure~\ref{fig.reductants}(b).
Therefore, \VCo(1/2/3)-\WO(1) has higher energy than \VCo(2/3/5)-\WO(9) by 0.47 eV;
as a result, it cannot be thermodynamically synthesized
because of its synthesis temperature (368 K)~\cite{2017LIN}.
The other two localized structures, (2-1) \VCo(3/5/6)-\WO(7)
and (2-2) \VCo(2/3/4)-\WO(6), have much higher energies
than the most stable structure, (1-1) \VCo(2/3/5)-\WO(9).
This can be attributed to the fact that Coulomb repulsions between
more localized \VCo{} pairs are greater than those between delocalized ones,
irrespective of H-bonds.
%--------------------------------
\begin{table}[]
  \begin{center}
    \caption{
      Four \VCo{} patterns in Figure~\ref{fig.reductants}
      and the corresponding most stable \WO{} sites,
      together with their relative energies ($\Delta E$).
      $\Delta E$ is set to be zero for \VCo(2/3/5)-\WO(9).
    }
    \begin{tabular}{cccc}
      Category & V$_{\rm{Co}}$($i/j/k$)&W$_{\rm{O}}$($l$)&$\Delta E$/eV \\ \hline
      1-1 &  2/3/5 & 9 & +0.00\\ 
      1-2 &  1/2/3 & 1 & +0.47\\
      2-1 &  3/5/6 & 7 & +0.89\\
      2-2 &  2/3/4 & 6 & +5.42\\ 
      \hline
      \label{tab.opt_hyd}
    \end{tabular}
  \end{center}
\end{table}
%--------------------------------
%--------------------------------
%--------------------------------

\vspace{2mm}
Herein, we investigate the relative stability
of \VCo(2/3/5) with respect to different sites of H$_2$O,
\WO($l$)~($l=1\sim 9$), listed in Table~\ref{tab.water}.
The three structures, \VCo(2/3/5)-\WO(6)/-\WO(1)/-\WO(5),
are more stable than (1-2) \VCo(1/2/3)-\WO(1),
but less stable than (1-1) \VCo(2/3/5)-\WO(9).
\VCo(2/3/5)-\WO(6)/-\WO(1)/-\WO(5)
have higher energies than \VCo(2/3/5)-\WO(9)
by $0.15/0.32/0.42$ eV, respectively.
Similar to \VCo(1/2/3)-\WO(1),
these metastable phases are not synthesized thermodynamically
due to their synthesis temperatures.
Three delocalized \VCo{} patterns,
\VCo(2/3/5)-\WO(2)/-\WO(4)/-\WO(8),
are less stable than the localized \VCo(3/5/6)-\WO(7) due
to the presence of highly sterically
hindered structural water sites.
Thus, we conclude that the CCH reductant has
the \VCo(2/3/5)-\WO(9) structure,
which is thermodynamically synthesizable.
Its structural stability can be attributed to
the abovementioned H-bonds.
%--------------------------------
\begin{table}[]
  \begin{center}
    \caption{
      Relative stability of \VCo(2/3/5)
      with respect to different sites of
      H$_2$O, \WO($l$)~($l=1\sim 9$).
      The relative energy $\Delta E$ is given in eV,
      setting the most stable \VCo(2/3/5)-\WO(9) pattern
      to be zero.
    }
\begin{tabular}{cc}
  W$_{\rm{O}}$($l$)&$\Delta E$/eV \\ \hline
   9 & +0.00\\ 
   6 & +0.15\\
   1 & +0.32\\
   5 & +0.42\\
   3 & +0.51\\
   7 & +0.68\\
   8 & +0.96\\
   2 & +1.37\\
   4 & +5.47\\
  \hline
  \label{tab.water}
\end{tabular}
  \end{center}
\end{table}
%--------------------------------

%sssssssssssssssssssssssssssssssssssssssssssssssssssss
\subsection{Identification of oxidant}
\label{oxidant}
%sssssssssssssssssssssssssssssssssssssssssssssssssssss
%--------------------------------
\begin{table}[]
  \begin{center}
  \caption{
  DFT-evaluated EMFs required for the deprotonation reactions
      ($V_\mathrm{dep}^\mathrm{DFT}$) at deprotonation sites, H$_\mathrm{dep}$,
      (see Figure~\ref{fig.hyd_position}).
      $V_\mathrm{dep}$ is given in unit of V (vs.SCE). 
    }
    \vspace{2mm}
    \begin{tabular}{ccc}
      H$_{\rm{dep}}$&$V_{\mathrm{dep}}^\mathrm{DFT}$(vs.SCE)/V \\ \hline
      7 & 3.05\\
      10& 3.08\\
      3 & 3.08\\
      2 & 3.26\\
      1 & 3.27\\
      8 & 3.30\\
      6 & 3.34\\
      9 & 3.52\\
      4 & 4.06\\
      5 & 4.89\\ \hline
      \label{tab.del_hyd}
    \end{tabular}
  \end{center}
\end{table}
%------------------
To investigate the electrochemical reactivity of CCH,
we obtained oxidant structures by removing H atoms
from the reductant structure, \VCo(2/3/5)-\WO(9),
where the deprotonation sites (H$_\mathrm{dep}$)
are numbered as $\mathrm{H}_1 \sim \mathrm{H}_{10}$
in Figure~\ref{fig.hyd_position}.
For computing the oxidant and reductant energies
through the DFT simulations,
Eq.~\eqref{eq:V} is applied to evaluate
their DFT-evaluated electromotive forces (EMFs) of
deprotonation reactions ($V_\mathrm{dep}^\mathrm{DFT}$)
occurring inside CCH crystals,
which are listed in Table~\ref{tab.del_hyd}.
According to these computational results,
all the deprotonation reactions require
$V_\mathrm{dep}^\mathrm{DFT} \ge 3$ V.
Compared with the computed EMFs,
the experimental EMFs ($V_\mathrm{dep}^\mathrm{exp}$)
are estimated to be less than $+0.6$~V
because the CV measurements for the CCH capacitors
report the potential window of $V_\mathrm{dep}^\mathrm{exp} < +0.6$ V
when the deprotonation reaction occurs.

\vspace{2mm}
The difference between $V_\mathrm{dep}^\mathrm{DFT}$ and $V_\mathrm{dep}^\mathrm{exp}$
implies that the deprotonation reaction occurs not inside
the CCH crystal, but on the surface.
Note that $V_\mathrm{dep}^\mathrm{DFT}$ approximates to
a definite EMF ($V_\mathrm{dep}$) because
$V_\mathrm{dep}$ should be defined as
the energy difference between the reductant and transition state
in the electrochemical reaction (i.e., activation energy),
whereas $V_\mathrm{dep}^\mathrm{DFT}$ is just the energy difference
between the reductant (before reaction)
and oxidant (after reaction) energies, according to Eq.~\eqref{eq:V}.
However, since $V_\mathrm{dep}^\mathrm{DFT}$ necessarily underestimates $V_\mathrm{dep}$,
the present $V_\mathrm{dep}^\mathrm{DFT}$ values are sufficient 
to prove whether or not the deprotonation reaction occurs
inside the CCH crystal.
Actually, it is quite hard to perform an exhaustive first-principles
transition state calculations~\cite{2020TOM} for all the H$_\mathrm{dep}$ sites
due to their computational costs.

\vspace{2mm}
The deprotonation inside the CCH bulk necessitates breaking both the H-bond and
the O-H covalent bond, which involves high energy of an eV order,
while the electrochemical reaction in actual CCH electrodes occurs
at less than $+0.6$ V(vs. SCE).
In other words, the CCH bulk itself
is stable against an electrochemical reaction.
The electrochemical stability in the CCH bulk
can be attributed to the H-bonds formed by all the H atoms
with the neighboring O atoms, as shown in Figure~\ref{fig.hyd_position}.

\vspace{2mm}
The computational results shown here suggest 
that a single phase of pristine CCH electrode has
a low capacitance.~\cite{2021SHU}
In contrast, some CCH-based electrodes have been reported
to exhibit high capacity.~\cite{2017LIN, 2021LIU}
The difference between high- and low-capacity CCH electrodes
can be attributed to their different morphology and constituent phases.
According to the experimental characterization,~\cite{2017LIN}
the high-capacity CCH electrodes consist of
not only a pristine CCH phase but also hydroxide (Co(OH)$_2$) phase.
Thus, we investigate the crystal growth mechanism of CCH
based on XRD analysis, which is the key to understand
how actual CCH electrodes are synthesized.

%%------------------
\subsection{Crystal growth designated by XRD peaks}
%%------------------
In terms of capacity,
there are two types of CCH-based electrodes:
CCH-based electrodes with high capacity~\cite{2017LIN}
and low capacity~\cite{2015GHO,2021SHU}
are synthesized with NO$_3^{-}$ and Cl$^-$ counterions for cobalt ions
(NA- and CA-CCH), respectively.
They have different morphologies that arise from
different crystal growth directions and degrees of crystallinity.
Differences in the crystal growth directions and degrees of crystallinity for CCH
can be explained from observed XRD patterns.

\vspace{2mm}
In fact, previous studies characterized
XRD patterns of the CCH-based electrodes,
but their XRD analyses were based on an incorrect lattice ansatz,
an ortholombic lattice.~\cite{1992POR}
The correct lattice of CCH has been recently
found to be a hexagonal lattice.~\cite{2019BHO}
Note that the crystal growth direction of a pristine CCH crystal
has been found to be $c$-axis.
Thus, the XRD patterns of the actual CCH electrodes should be
reassigned into Miller indices.

\begin{figure}[]
  \centering
  \includegraphics[width=0.99\hsize]{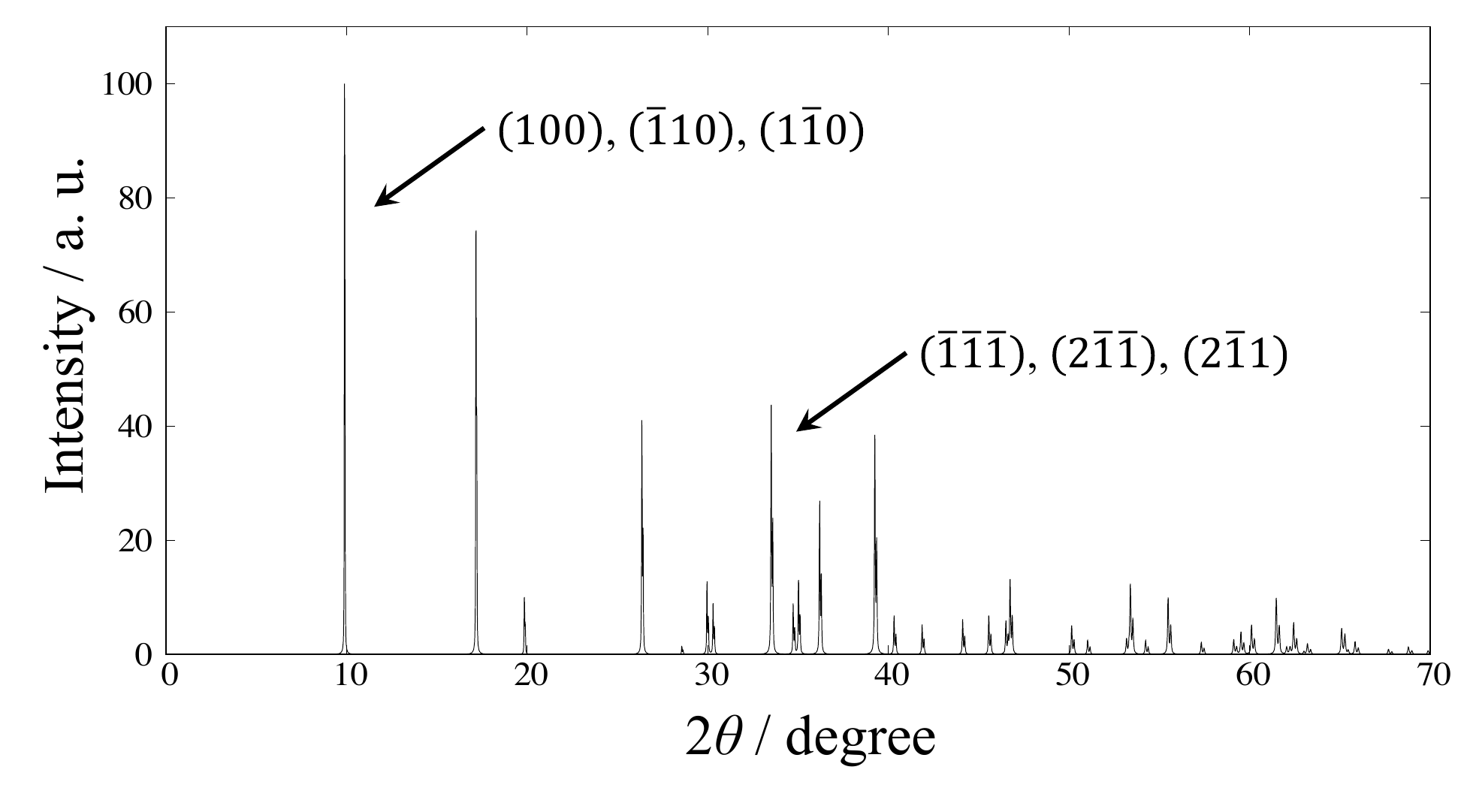}
  \caption{
    \label{fig.XRD}
    XRD peak pattern obtained from the XRD simulation
    for the computed hexagonal CCH crystal structure (3D NA-CCH).
    The peak at $2\theta=9.885$ and $33.515$ includes
    a plane set $\{(100), (\bar{1}10), (1\bar{1}0)\}$
    and $\{(\bar{1}\bar{1}\bar{1}), (2\bar{1}\bar{1}), (2\bar{1}1)\}$,
    respectively.
    They were incorrectly assigned in previous studies,
    because the crystal structure was assumed to be not
    hexagonal, but orthorhombic.
  }
\end{figure}

\vspace{2mm}
In order to correctly assign XRD peaks to Miller indices, 
we performed the XRD simulation based on
the correct CCH crystal structure obtained computationally,
where its unit cell came from NA-CCH~\cite{2019BHO}
and occupancy of structure water site was set to be $1.0$.
Note that our assignment of lattice planes is different
from the previous one,~\cite{2003XU}
because the correct CCH structure is different from
previously characterized ones.
Then, we examined which plane contribute the morphology difference of
NA- and CA-CCH electrodes from the XRD simulation results.
It is well known that a XRD peak position $(2\theta)$
identifies certain lattice planes,
whereas the relative peak intensity shows
an amount of the planes formed in a sample material.
Figure~\ref{fig.XRD} shows the XRD peaks
in terms of relative intensity
where the strongest value is set to be 100.
%--------------------------------
%--------------------------------

\par
\vspace{2mm}
Here we focus on the two peaks at $2\theta=9.885$ and $33.515$;
the former is related to the crystal growth direction almost parallel to
$c$ axis, whereas the latter to that toward $ab$-plane, as shown below.
Note that each of the peaks consists of several lattice planes
with different intensities,
but our concern is only main peaks having their intensities
more than 20\% of the strongest intensity.
Eventually, the peaks at $2\theta=9.885$ and $33.515$ correspond to
sets of lattice planes (1) $\{(100), (\bar{1}10), (1\bar{1}0)\}$ and
(2) $\{(\bar{1}\bar{1}\bar{1}), (2\bar{1}\bar{1}), (2\bar{1}1)\}$, respectively;
their structures are given in Supporting Information (SI).
Note that only representative lattice planes appear in Figure~\ref{fig.XRD},
whereas their equivalent planes are omitted.
(1) Looking at the former planes, we can see that
H-bonds (encompassed with green rectangular)
are formed on $ab$-planes connecting the $(100)$ planes (Figure S-1).
These H-bonds can promote the 2D crystal growth
($a$- and $b$-axis directions).
The above finding is similar to the $(1\bar{1}0)$ and $(\bar{1}10)$ planes.
(2) For the latter, H-bonds are formed so as to
connect the $(\bar{1}\bar{1}\bar{1})$ planes,
whose direction is almost parallel to $c$-axis (Figure S-2).
These H-bonds can work for the 1D crystal growth
($c$-axis direction),
which can be called ``H-bond-driven 1D crystal growth''.
This holds for the $(2\bar{1}\bar{1})$ and $(2\bar{1}1)$ planes as well.
Thus, a comparison between $I(9.885)$ and $I(33.515)$
can tell us about the crystal growth direction.

\par
\vspace{2mm}
The NA- and CA-CCH electrodes have different surfaces with specific lattice planes,
which is caused by the difference in their different crystal growth directions.
To investigate their difference, 
we make a comparison of their intensities.
For that purpose, we adopt a ratio of the peak intensity at $2\theta = 33.515$ to that at $2\theta=9.885$:
%------------------
\begin{equation}
  R = \frac{I(33.515)}{I(9.885)}.
\end{equation}
%------------------
Recall that the simulated XRD peaks are computed for an isotropic CCH bulk.
In contrast, anisotropic CCH electrodes have different $I(2\theta)$ values from the bulk,
though they have the same $2\theta$ as the bulk.
Thus, the NA-CCH and CA-CCH electrodes have different $R$ values from the bulk.
Table~\ref{tab:peakRatio} lists 
$I(2\theta)$ at $2\theta = 9.885$ and $33.515$ and $R$ for the bulk
and the NA- and CA-CCH electrodes (obtained from Figure 1 (b)
in Ref.~\cite{2003XU}).
It is found that 
the NA- and CA-CCH electrodes have $R \simeq 1.7$ and $\simeq 0.44$,
respectively.
As we mentioned, relatively stronger peaks mean
more amount of the corresponding planes than others
formed in a material.
Thus, the plane set $\{(\bar{1}\bar{1}\bar{1}), (2\bar{1}\bar{1}), (2\bar{1}1)\}$
is formed more than the set $\{(100), (\bar{1}10), (1\bar{1}0)\}$ in NA-CCH.
This means that NA-CCH exhibits the 1D crystal growth
toward $c$-axis.
In contrast, $\{(100), (\bar{1}10), (1\bar{1}0)\}$ is formed more than
$\{(\bar{1}\bar{1}\bar{1}), (2\bar{1}\bar{1}), (2\bar{1}1)\}$ in CA-CCH,
indicating that
the crystal growth direction of CA-CCH is more 2D than NA-CCH.
Here we also see that
the $R$ value of CA-CCH is almost same as that of the bulk NA-CCH.
This indicates that in CA-CCH,
the 1D crystal growth toward $c$-axis is
balanced with the 2D one inside $ab$-planes.
In other words, the CA-CCH structure is 3D,
similar to the bulk NA-CCH structure.
To sum up, the NA-CCH electrode is 1D,
whereas the CA-CCH one is 3D,
which can be attributed to the H-bonds formed therein.

%------------------
\begin{table*}[]
  \begin{center}
    \caption{
      Peak intensities $I(2\theta)$ at $2\theta = 9.885$ and $33.515$,
      and the corresponding intensity ratio $R$
      for the NA- and CA-CCH electrode materials and the bulk.
      The $I(2\theta)$ values for the NA- and CA-CCH materials
      are obtained from Reference~\cite{2003XU},
      whereas the corresponding bulk NA-CCH values come from
      the present XRD simulations.
    }
    \begin{tabular}{cccc}\hline
       & $I(9.885)$ & $I(33.515)$ & $R$ \\
      \hline
      NA-CCH~\cite{2003XU}& $\simeq$ 60 & $\simeq$ 100 & $\simeq$ 1.7 \\ 
      CA-CCH~\cite{2003XU}& $\simeq$ 80 & $\simeq$ 35 & $\simeq$ 0.44 \\ 
      bulk NA-CCH(present study) & 100 & 44 & 0.44 \\
      \hline
      \label{tab:peakRatio}
    \end{tabular}
  \end{center}
\end{table*}
%--------------------------------

\vspace{2mm}
We have investigated why the directions of H-bonds in NA-CCH
are different from those in CA-CCH.
NO$_3^{-}$ and CO$_3^{2-}$ are used as cobalt counter anions
to synthesize NA- and CA-CCH, respectively.
Their difference in reactants
gives rise to that in occupancy of structural waters.~\cite{2003XU}
It has been observed experimentally that NA-/CA-CCH
with a higher/lower H$_2$O occupancy
grow one/two-dimensionally,
implying that the structural waters inhibit
the 2D crystal growth.~\cite{2003XU}
As shown in Figure S1, a structural water
(encompassed with an orange ellipse) forms
two H-bonds between the $(\bar{1}\bar{1}\bar{1})$ planes.
The $c$-axis H-bond involving the structural water
is considered to promote
the one-dimensional crystal growth of NA-CCH.
In contrast, the structural water in CCH
can be regarded as a growth inhibitor in the 2D direction.

\vspace{2mm}
In addition to the peak intensities,
the peak broadening provides an insight into
the capacitance difference between the NA- and CA-CCH electrodes.
NA-CCH was found to have broad peaks in the overall angle range.~\cite{2003XU}
The presence of these broad peaks can be attributed to the size of crystals and
defects in the NA-CCH electrodes.
The Co(OH)$_2$ phase is present in the NA-CCH electrode.~\cite{2017LIN}
Since the NA-CCH crystal growth direction is $c$-axis,~\cite{2019BHO}
Co(OH)$_2$ layers are anisotropically stacked
along the $c$-axis direction,
when synthesizing the NA-CCH electrode with CCH precursors.
The Co(OH)$_2$ is considered to be the $\beta$ phase:
it has two polymorphs, $\alpha/\beta$-Co(OH)$_2$,
where the $\alpha$ phase is metastable
and transformed easily into the $\beta$ phase.~\cite{2002GAU}
This indicates that
$\beta$-Co(OH)$_2$ rather than $\alpha$-Co(OH)$_2$
is synthesized using the CCH precursor.
Thus, we focus on the crystal growth and
electrochemical reaction of the pristine $\beta$-Co(OH)$_2$ phase as well.
The pristine $\beta$-Co(OH)$_2$ crystal has pore layers,
and hence OH$^-$ ions enter into the pore layers,
thereby leading to the deprotonation reaction in the crystal.
In contrast, it has been observed from TEM images
that plate surfaces contact with interlayers during the crystal growth
to form the aggregated bulk electrode, {\it i.e.},
thick plate structures of $\beta$-Co(OH)$_2$ aggregate into
the $\beta$-Co(OH)$_2$ bulk electrode.~\cite{2017DEN}
The contact site of the interlayer and plate surface
can be prevent OH$^-$ ions from
entering into the interlayer where the deprotonation reaction occurs.
The Co(OH)$_2$ phase is formed from the crystal surface
by decarboxylation reaction between CCH and OH$^-$.
The pore layers of deprotonation sites in the $\beta$-Co(OH)$_2$
are oriented to electrolyte.
Eventually, more anisotropic $\beta$-Co(OH)$_2$ is synthesized than
pristine $\beta$-Co(OH)$_2$.

%%%%%%%%%%%%%%%%%%%%%%%%%%%%%%%%%%
\section{Discussion}
\label{discussion}
%sssssssssssssssssssssssssssssssssssssssssssssssssssss
%%------------------
\subsection{Morphology-driven storage \\properties}
%%------------------
Table~\ref{tab:perform} lists storage properties
(capacity and cycle life) of
high- and low-capacity CCH electrodes,
NA-CCH~\cite{2017LIN} and CA-CCH~\cite{2015GHO,2021SHU},
as well as pristine $\alpha/\beta$-Co(OH)$_2$ electrodes
for comparison,
which has been reported in previous studies.
Their differences can be interpreted in terms of
their morphologies, as shown below.
%------------------
\begin{table*}[t]
  \begin{center}
    \caption{Capacitance and cycle characteristics of Co-based
      electrode materials, associated with morphologies:
      CFP, NF, SS, and PN indicates
      substrates of carbon fiber paper, nickel foam,
      stainless steel, and porous Ni, respectively.
    }
    \begin{tabular}{cccc}\hline
      electrode & morphology & capacity & cycle life \\
      & & @[electron density] & @[charge density]/(cycles) \\
      & & /Fg$^{-1}$@Ag$^{-1}$ & /\%@Ag$^{-1}$ \\ \hline
      CCH@NF~\cite{2021LIU}& umbrella-like NWA & 1227[@2.25] & 93.7[@4.0](10000) \\ 
      CCH@NF~\cite{2017LIN}& umbrella-like NWA & 1381[@2.0]  & 93.5[@40.0](5000) \\ 
      CCH@SS~\cite{2021SHU}& nanorod-assembled & 165.6[@0.1] &$\sim$100.0[@0.5](1000) \\
      $\alpha$-Co(OH)$_2$~\cite{2010KON}    & porous film & 1473[@2.0]  & 88.0[@6](1000) \\      
      $\beta$-Co(OH)$_2$@CFP~\cite{2017DEN} & nano plate & 800[@2.0]  & 95.7[@2](8000) \\
      
      \hline
      \label{tab:perform}
    \end{tabular}
  \end{center}
\end{table*}
%--------------------------------

\vspace{2mm}
First, we investigate the CA-CCH electrodes.
As mentioned previously,
the CA-CCH electrodes are composed of single crystals,
where the Co(OH)$_2$ phases hardly exist inside the crystal.
In addition, the CA-CCH electrodes have less defect with smaller
surface areas (relative to bulk region) than the NA-CCH.
This implies that OH$^{-}$ ions in the electrolytic solution
are hindered from diffusing into the inside region in the CCH crystal.
In other words, the OH$^-$ ion diffusions just occur in restricted surface areas.
These conjectures would give rise to the low capacitance of 165 F/g
in the CCH electrode aggregated on a stainless-steel substrate
forming nanorod-assembled hierarchical structures~\cite{2021SHU},
as shown in Table~\ref{tab:perform}.
This assembled structure consists of
1-D aggregated rod,
whereas each rod unit is two-dimensionally grown.
The 2-D morphology causes
the deprotonation reaction to occur only around the surface regions; therefore,
the reaction become reversible,
leading to the ``perfect'' cycle life,
as shown in Table~\ref{tab:perform}.

\vspace{2mm}
In contrast, the NA-CCH electrodes have significantly different
morphologies from the CA-CCH ones:
NA-CCH is composed of not only CCH but also Co(OH)$_2$ phases,
i.e., multiphase, with small crystal sizes and defects,
compared to NA-CCH.~\cite{2003XU,2017LIN,2021LIU}
The smaller crystal sizes lead to the wider
surface areas compared to the bulk region,
increasing the diffusion sites of OH$^-$ ions on the surface
and hence forming the Co(OH)$_2$ phase.
The surface defects and grain boundaries
also enhance the OH$^-$ diffusions inside the single crystals
and on the single-crystal surfaces located in the polycrystals, respectively.
Specifically, grain boundaries may promote
the formation of the Co(OH)$_2$ phase.
Eventually, the high capacity of NA-CCH
can be attributed to the above characteristic morphologies.
The 1-D morphology enhances
the deprotonation reaction occurring at the Co(OH)$_2$ phases.
Therefore, the NA-CCH morphology leads to higher capacity, but lower life cycle,
compared to those of CA-CCH.
However, NA-CCH partially contains the CCH phase,
which make the electrode more stable than
the pristine Co(OH)$_2$ phase.
This is the reason why NA-CCH has longer
life cycles ($\sim 93$\%{}) than the pristine $\alpha$-Co(OH)$_2$ electrode.
In contrast, the pristine $\beta$-Co(OH)$_2$ electrode
has longer life cycles ($\sim 96$ \%)
than the NA-CCH electrodes ($\sim 94$\%),
but in turn has a significantly lower capacity ($800$ F/g).
In general, there exist a counterbalance
between capacity and life cycle.

%%------------------
\subsection{Further details for NA-CCH}
%%------------------
Within the framework of the NA-CCH electrodes,
their morphologies significantly vary
depending on their growth times,
which affects their storage performance.
For instance, as per Liu et al.,~\cite{2017LIN}
the CCH precursors were grown on NF substrates
and the morphologies of the NA-CCH electrodes
controlled the growth times of 6 and 10 h
(designated as NA-CCH@NF-6h and NA-CCH@NF-10h, respectively).
Both the electrodes have ewire structures
having larger surface areas perpendicular to their substrates,
compared to the CA-CCH electrodes.
However, upon comparing them, 
a remarkable difference in
the wire orientations was found relative to the substrate.

\vspace{2mm}
NA-CCH@NF-10h has an umbrella-like nanowire structure,
which corresponds to that listed in Table~\ref{tab:perform}.
In contrast, NA-CCH@NF-6h has a nanoneedle structure
oriented perpendicular to the substrate.
According to experimental observations based on XRD and TEM,
the NA-CCH@NF-6h structure is more uniaxially-anisotropic
than the NA-CCH@NF-10h one,
hence the former has wider surface areas on the sides
perpendicular to the substrate than the latter.

\vspace{2mm}
Their structural/morphological difference results in the difference in the storage performance.
Actually, NA-CCH@NF-6h has a higher capacity (1548 F/g(@2 A/g))
than NA-CCH@NF-10h (1381 F/g(@2 A/g))
at the initial stage of charge-discharge cycle.
This can be explained as follows:
NA-CCH@NF-6h and NA-CCH@NF-10h are almost identical to each other
in terms of the broadness of the XRD peak patterns.
This implies that they have almost same defect structures.
Therefore, the capacity difference between NA-CCH@NF-6h
and NA-CCH@NF-10h can be attributed to their difference
in the surface areas; during the initial cycles,
the OH$^-$ ions react with the CCH surface
and further diffuse into the inside crystal.
This reaction in NA-CCH@NF-6h enhances the capacity,
but rapidly degrades the cycle stability,
giving rise to the deformation of the structure.
Because of its 1-D morphology,
the unstable structure in NA-CCH@NF-6h
can be attributed to the less hydrogen bonds
and chemical bonds with CO$_3^{2-}$
than the pristine CCH crystal (Figure~\ref{fig.structure} (a)).
In the middle stage of the charge-discharge cycle,
the structural change inhibits the OH$^-$ ions
from diffusing into the interlayers inside the crystal.
Finally, the NA-CCH@NF-6h electrode melts down,
losing the storage performance.

\vspace{2mm}
Recently, the storage performance has been reported
for a CCH electrode similar to the NA-CCH@NF-10h one,
exhibiting a high capacity of 1227 F/g(@2.25 A/g)
and high cycle stability of 93.7 \% (@4 A/g)
after 10,000 cycles.~\cite{2021LIU}
The CCH phase contributes to the cycle stability
due to its hydrogen and chemical bonds,
whereas the Co(OH)$_2$ phase enhances the capacity
albeit the less stability.
The above high performance was achieved by counterbalancing
the two phases.
This study has highlighted the possibility of regulating
the ratio of the CCH phase versus the Co(OH)$_2$ phase
by controlling the reaction surface area.
%%%%%%%%%%%%%%%%%%%%%%%%%%%%%%%%%%
\section{Conclusion}
\label{conclusion}
We performed exhaustive first-principles simulations
to identify a hexagonal CCH structure,
especially for atomic positions of cobalt vacancies and hydrogen atoms.
The resultant structure is a structural model of reductant
in the electrochemical reaction of CCH-based electrodes.
This was used to explore the oxidant structure.
The reductant and oxidant structures help us evaluate the EMF of the deprotonation reaction.
The computed EMF value of $\sim 3$ V (vs. SCE) was much larger than
that expected from the experimental potential window of $< 0.6$ V (vs. SCE).
This indicates that the deprotonation reaction never occurs
inside the CCH bulk, and therefore,
the reaction sites are restricted to the surface regions
in the pristine CCH electrode.
The detailed information about the atomic positions revealed that
the chemical and hydrogen bonds are formed between the
$\{(\bar{1}\bar{1}\bar{1}), (2\bar{1}\bar{1}), (2\bar{1}1)\}$ planes,
which prevent the hydroxide (OH$^-$) ions diffusing into the inside crystal.
We also performed the XRD simulation to assign XRD peaks to
the Miller indices based on the computed structure with 
the correct hexagonal lattice.
The reassigned Miller indices revised the crystal growth direction
to be specified by the $\{(\bar{1}\bar{1}\bar{1}), (2\bar{1}\bar{1}), (2\bar{1}1)\}$ planes.

\vspace{2mm}
According to the above computational findings,
we investigated how some CCH electrodes achieve
a better storage performance than pristine
Co(OH)$_2$ electrodes.
The revised crystal growth direction clarified
the morphologies formed in the high-capacity CCH electrode
synthesized with the NO$_3^{2-}$ counterion.
The results help us understand the importance of both
counterbalancing the coexisting CCH and Co(OH)$_2$ phases
and controlling their morphologies
(e.g., forming an umbrella-like morphology)
for achieving the best storage performance.

%%%%%%%%%%%%%%%%%%%%%%%%%%%%%%%%%%
\section{Acknowledgments}
The computations in this work have been performed 
using the facilities of Research Center for Advanced Computing Infrastructure (RCACI) at JAIST. 
K.O. is grateful for the final support from JST SPRING (JPMJSP2102).
R.M. is grateful for financial supports from MEXT-KAKENHI (22H05146, 21K03400, and 19H04692), 
from the Air Force Office of Scientific Research (AFOSR-AOARD/FA2386-17-1-4049; FA2386-19-1-4015).
K.H. is grateful for financial support from 
MEXT-KAKENHI, Japan (JP19K05029, JP21K03400, JP21H01998, and JP22H02170), 
and the Air Force Office of Scientific Research, United States (Award Numbers: FA2386-20-1-4036). 

\bibliography{references.bib}

\begin{thebibliography}{29}%
\makeatletter
\providecommand \@ifxundefined [1]{%
 \@ifx{#1\undefined}
}%
\providecommand \@ifnum [1]{%
 \ifnum #1\expandafter \@firstoftwo
 \else \expandafter \@secondoftwo
 \fi
}%
\providecommand \@ifx [1]{%
 \ifx #1\expandafter \@firstoftwo
 \else \expandafter \@secondoftwo
 \fi
}%
\providecommand \natexlab [1]{#1}%
\providecommand \enquote  [1]{``#1''}%
\providecommand \bibnamefont  [1]{#1}%
\providecommand \bibfnamefont [1]{#1}%
\providecommand \citenamefont [1]{#1}%
\providecommand \href@noop [0]{\@secondoftwo}%
\providecommand \href [0]{\begingroup \@sanitize@url \@href}%
\providecommand \@href[1]{\@@startlink{#1}\@@href}%
\providecommand \@@href[1]{\endgroup#1\@@endlink}%
\providecommand \@sanitize@url [0]{\catcode `\\12\catcode `\$12\catcode
  `\&12\catcode `\#12\catcode `\^12\catcode `\_12\catcode `\%12\relax}%
\providecommand \@@startlink[1]{}%
\providecommand \@@endlink[0]{}%
\providecommand \url  [0]{\begingroup\@sanitize@url \@url }%
\providecommand \@url [1]{\endgroup\@href {#1}{\urlprefix }}%
\providecommand \urlprefix  [0]{URL }%
\providecommand \Eprint [0]{\href }%
\providecommand \doibase [0]{http://dx.doi.org/}%
\providecommand \selectlanguage [0]{\@gobble}%
\providecommand \bibinfo  [0]{\@secondoftwo}%
\providecommand \bibfield  [0]{\@secondoftwo}%
\providecommand \translation [1]{[#1]}%
\providecommand \BibitemOpen [0]{}%
\providecommand \bibitemStop [0]{}%
\providecommand \bibitemNoStop [0]{.\EOS\space}%
\providecommand \EOS [0]{\spacefactor3000\relax}%
\providecommand \BibitemShut  [1]{\csname bibitem#1\endcsname}%
\let\auto@bib@innerbib\@empty
%</preamble>
\bibitem [{\citenamefont {Fleischmann}\ \emph {et~al.}(2020)\citenamefont
  {Fleischmann}, \citenamefont {Mitchell}, \citenamefont {Wang}, \citenamefont
  {Zhan}, \citenamefont {en~Jiang}, \citenamefont {Presser},\ and\
  \citenamefont {Augustyn}}]{2020FLE}%
  \BibitemOpen
  \bibfield  {author} {\bibinfo {author} {\bibfnamefont {S.}~\bibnamefont
  {Fleischmann}}, \bibinfo {author} {\bibfnamefont {J.~B.}\ \bibnamefont
  {Mitchell}}, \bibinfo {author} {\bibfnamefont {R.}~\bibnamefont {Wang}},
  \bibinfo {author} {\bibfnamefont {C.}~\bibnamefont {Zhan}}, \bibinfo {author}
  {\bibfnamefont {D.}~\bibnamefont {en~Jiang}}, \bibinfo {author}
  {\bibfnamefont {V.}~\bibnamefont {Presser}}, \ and\ \bibinfo {author}
  {\bibfnamefont {V.}~\bibnamefont {Augustyn}},\ }\href {\doibase
  10.1021/acs.chemrev.0c00170} {\bibfield  {journal} {\bibinfo  {journal}
  {Chemical Reviews}\ }\textbf {\bibinfo {volume} {120}},\ \bibinfo {pages}
  {6738} (\bibinfo {year} {2020})}\BibitemShut {NoStop}%
\bibitem [{\citenamefont {Liang}\ \emph {et~al.}(2021)\citenamefont {Liang},
  \citenamefont {Du}, \citenamefont {Xiao}, \citenamefont {Cheng},
  \citenamefont {Yuan}, \citenamefont {Chen}, \citenamefont {Yuan},\ and\
  \citenamefont {Chen}}]{2021LIA}%
  \BibitemOpen
  \bibfield  {author} {\bibinfo {author} {\bibfnamefont {R.}~\bibnamefont
  {Liang}}, \bibinfo {author} {\bibfnamefont {Y.}~\bibnamefont {Du}}, \bibinfo
  {author} {\bibfnamefont {P.}~\bibnamefont {Xiao}}, \bibinfo {author}
  {\bibfnamefont {J.}~\bibnamefont {Cheng}}, \bibinfo {author} {\bibfnamefont
  {S.}~\bibnamefont {Yuan}}, \bibinfo {author} {\bibfnamefont {Y.}~\bibnamefont
  {Chen}}, \bibinfo {author} {\bibfnamefont {J.}~\bibnamefont {Yuan}}, \ and\
  \bibinfo {author} {\bibfnamefont {J.}~\bibnamefont {Chen}},\ }\href {\doibase
  10.3390/nano11051248} {\ \textbf {\bibinfo {volume} {11}},\ \bibinfo {pages}
  {1248} (\bibinfo {year} {2021})}\BibitemShut {NoStop}%
\bibitem [{\citenamefont {Lin}\ \emph {et~al.}(2017)\citenamefont {Lin},
  \citenamefont {Li}, \citenamefont {Musharavati}, \citenamefont {Zalnezhad},
  \citenamefont {Bae}, \citenamefont {Cho},\ and\ \citenamefont
  {Hui}}]{2017LIN}%
  \BibitemOpen
  \bibfield  {author} {\bibinfo {author} {\bibfnamefont {X.}~\bibnamefont
  {Lin}}, \bibinfo {author} {\bibfnamefont {H.}~\bibnamefont {Li}}, \bibinfo
  {author} {\bibfnamefont {F.}~\bibnamefont {Musharavati}}, \bibinfo {author}
  {\bibfnamefont {E.}~\bibnamefont {Zalnezhad}}, \bibinfo {author}
  {\bibfnamefont {S.}~\bibnamefont {Bae}}, \bibinfo {author} {\bibfnamefont
  {B.-Y.}\ \bibnamefont {Cho}}, \ and\ \bibinfo {author} {\bibfnamefont
  {O.~K.~S.}\ \bibnamefont {Hui}},\ }\href {\doibase 10.1039/c7ra09050a}
  {\bibfield  {journal} {\bibinfo  {journal} {{RSC} Adv.}\ }\textbf {\bibinfo
  {volume} {7}},\ \bibinfo {pages} {46925} (\bibinfo {year}
  {2017})}\BibitemShut {NoStop}%
\bibitem [{\citenamefont {Shu}\ \emph {et~al.}(2021)\citenamefont {Shu},
  \citenamefont {Liang}, \citenamefont {Zhang},\ and\ \citenamefont
  {Fang}}]{2021SHU}%
  \BibitemOpen
  \bibfield  {author} {\bibinfo {author} {\bibfnamefont {C.}~\bibnamefont
  {Shu}}, \bibinfo {author} {\bibfnamefont {Y.}~\bibnamefont {Liang}}, \bibinfo
  {author} {\bibfnamefont {Z.}~\bibnamefont {Zhang}}, \ and\ \bibinfo {author}
  {\bibfnamefont {B.}~\bibnamefont {Fang}},\ }\href {\doibase
  10.1002/ejic.202100057} {\bibfield  {journal} {\bibinfo  {journal} {European
  Journal of Inorganic Chemistry}\ }\textbf {\bibinfo {volume} {2021}},\
  \bibinfo {pages} {1659} (\bibinfo {year} {2021})}\BibitemShut {NoStop}%
\bibitem [{\citenamefont {Deng}\ \emph {et~al.}(2009)\citenamefont {Deng},
  \citenamefont {Huang}, \citenamefont {Sun}, \citenamefont {Tsai},\ and\
  \citenamefont {Chang}}]{2009DEN}%
  \BibitemOpen
  \bibfield  {author} {\bibinfo {author} {\bibfnamefont {M.-J.}\ \bibnamefont
  {Deng}}, \bibinfo {author} {\bibfnamefont {F.-L.}\ \bibnamefont {Huang}},
  \bibinfo {author} {\bibfnamefont {I.-W.}\ \bibnamefont {Sun}}, \bibinfo
  {author} {\bibfnamefont {W.-T.}\ \bibnamefont {Tsai}}, \ and\ \bibinfo
  {author} {\bibfnamefont {J.-K.}\ \bibnamefont {Chang}},\ }\href {\doibase
  10.1088/0957-4484/20/17/175602} {\bibfield  {journal} {\bibinfo  {journal}
  {Nanotechnology}\ }\textbf {\bibinfo {volume} {20}},\ \bibinfo {pages}
  {175602} (\bibinfo {year} {2009})}\BibitemShut {NoStop}%
\bibitem [{\citenamefont {Mirzaeian}\ \emph {et~al.}(2020)\citenamefont
  {Mirzaeian}, \citenamefont {Akhanova}, \citenamefont {Gabdullin},
  \citenamefont {Kalkozova}, \citenamefont {Tulegenova}, \citenamefont
  {Nurbolat},\ and\ \citenamefont {Abdullin}}]{2020MIR}%
  \BibitemOpen
  \bibfield  {author} {\bibinfo {author} {\bibfnamefont {M.}~\bibnamefont
  {Mirzaeian}}, \bibinfo {author} {\bibfnamefont {N.}~\bibnamefont {Akhanova}},
  \bibinfo {author} {\bibfnamefont {M.}~\bibnamefont {Gabdullin}}, \bibinfo
  {author} {\bibfnamefont {Z.}~\bibnamefont {Kalkozova}}, \bibinfo {author}
  {\bibfnamefont {A.}~\bibnamefont {Tulegenova}}, \bibinfo {author}
  {\bibfnamefont {S.}~\bibnamefont {Nurbolat}}, \ and\ \bibinfo {author}
  {\bibfnamefont {K.}~\bibnamefont {Abdullin}},\ }\href {\doibase
  10.3390/en13195228} {\bibfield  {journal} {\bibinfo  {journal} {Energies}\
  }\textbf {\bibinfo {volume} {13}},\ \bibinfo {pages} {5228} (\bibinfo {year}
  {2020})}\BibitemShut {NoStop}%
\bibitem [{\citenamefont {Kong}\ \emph {et~al.}(2010)\citenamefont {Kong},
  \citenamefont {Liu}, \citenamefont {Lang}, \citenamefont {Liu}, \citenamefont
  {Luo},\ and\ \citenamefont {Kang}}]{2010KON}%
  \BibitemOpen
  \bibfield  {author} {\bibinfo {author} {\bibfnamefont {L.-B.}\ \bibnamefont
  {Kong}}, \bibinfo {author} {\bibfnamefont {M.-C.}\ \bibnamefont {Liu}},
  \bibinfo {author} {\bibfnamefont {J.-W.}\ \bibnamefont {Lang}}, \bibinfo
  {author} {\bibfnamefont {M.}~\bibnamefont {Liu}}, \bibinfo {author}
  {\bibfnamefont {Y.-C.}\ \bibnamefont {Luo}}, \ and\ \bibinfo {author}
  {\bibfnamefont {L.}~\bibnamefont {Kang}},\ }\href {\doibase
  10.1007/s10008-010-1125-6} {\ \textbf {\bibinfo {volume} {15}},\ \bibinfo
  {pages} {571} (\bibinfo {year} {2010})}\BibitemShut {NoStop}%
\bibitem [{\citenamefont {Deng}\ \emph {et~al.}(2017)\citenamefont {Deng},
  \citenamefont {Zhang}, \citenamefont {Arcelus}, \citenamefont {Kim},
  \citenamefont {Carrasco}, \citenamefont {Yoo}, \citenamefont {Zheng},
  \citenamefont {Wang}, \citenamefont {Tian}, \citenamefont {Zhang},
  \citenamefont {Cui},\ and\ \citenamefont {Rojo}}]{2017DEN}%
  \BibitemOpen
  \bibfield  {author} {\bibinfo {author} {\bibfnamefont {T.}~\bibnamefont
  {Deng}}, \bibinfo {author} {\bibfnamefont {W.}~\bibnamefont {Zhang}},
  \bibinfo {author} {\bibfnamefont {O.}~\bibnamefont {Arcelus}}, \bibinfo
  {author} {\bibfnamefont {J.-G.}\ \bibnamefont {Kim}}, \bibinfo {author}
  {\bibfnamefont {J.}~\bibnamefont {Carrasco}}, \bibinfo {author}
  {\bibfnamefont {S.~J.}\ \bibnamefont {Yoo}}, \bibinfo {author} {\bibfnamefont
  {W.}~\bibnamefont {Zheng}}, \bibinfo {author} {\bibfnamefont
  {J.}~\bibnamefont {Wang}}, \bibinfo {author} {\bibfnamefont {H.}~\bibnamefont
  {Tian}}, \bibinfo {author} {\bibfnamefont {H.}~\bibnamefont {Zhang}},
  \bibinfo {author} {\bibfnamefont {X.}~\bibnamefont {Cui}}, \ and\ \bibinfo
  {author} {\bibfnamefont {T.}~\bibnamefont {Rojo}},\ }\href {\doibase
  10.1038/ncomms15194} {\bibfield  {journal} {\bibinfo  {journal} {Nature
  Communications}\ }\textbf {\bibinfo {volume} {8}} (\bibinfo {year} {2017}),\
  10.1038/ncomms15194}\BibitemShut {NoStop}%
\bibitem [{\citenamefont {Ji}\ \emph {et~al.}(2015)\citenamefont {Ji},
  \citenamefont {Cheng}, \citenamefont {Yang}, \citenamefont {Jiang},
  \citenamefont {jie Jiang}, \citenamefont {Yang}, \citenamefont {Zhang},\ and\
  \citenamefont {Liu}}]{2015JI}%
  \BibitemOpen
  \bibfield  {author} {\bibinfo {author} {\bibfnamefont {X.}~\bibnamefont
  {Ji}}, \bibinfo {author} {\bibfnamefont {S.}~\bibnamefont {Cheng}}, \bibinfo
  {author} {\bibfnamefont {L.}~\bibnamefont {Yang}}, \bibinfo {author}
  {\bibfnamefont {Y.}~\bibnamefont {Jiang}}, \bibinfo {author} {\bibfnamefont
  {Z.}~\bibnamefont {jie Jiang}}, \bibinfo {author} {\bibfnamefont
  {C.}~\bibnamefont {Yang}}, \bibinfo {author} {\bibfnamefont {H.}~\bibnamefont
  {Zhang}}, \ and\ \bibinfo {author} {\bibfnamefont {M.}~\bibnamefont {Liu}},\
  }\href {\doibase 10.1016/j.nanoen.2014.11.064} {\bibfield  {journal}
  {\bibinfo  {journal} {Nano Energy}\ }\textbf {\bibinfo {volume} {11}},\
  \bibinfo {pages} {736} (\bibinfo {year} {2015})}\BibitemShut {NoStop}%
\bibitem [{\citenamefont {Choi}\ \emph {et~al.}(2013)\citenamefont {Choi},
  \citenamefont {Yang}, \citenamefont {Jung}, \citenamefont {Lee},
  \citenamefont {Kim}, \citenamefont {Park}, \citenamefont {Park},
  \citenamefont {Lee}, \citenamefont {Han},\ and\ \citenamefont
  {Huh}}]{2013CHO}%
  \BibitemOpen
  \bibfield  {author} {\bibinfo {author} {\bibfnamefont {B.~G.}\ \bibnamefont
  {Choi}}, \bibinfo {author} {\bibfnamefont {M.}~\bibnamefont {Yang}}, \bibinfo
  {author} {\bibfnamefont {S.~C.}\ \bibnamefont {Jung}}, \bibinfo {author}
  {\bibfnamefont {K.~G.}\ \bibnamefont {Lee}}, \bibinfo {author} {\bibfnamefont
  {J.-G.}\ \bibnamefont {Kim}}, \bibinfo {author} {\bibfnamefont
  {H.}~\bibnamefont {Park}}, \bibinfo {author} {\bibfnamefont {T.~J.}\
  \bibnamefont {Park}}, \bibinfo {author} {\bibfnamefont {S.~B.}\ \bibnamefont
  {Lee}}, \bibinfo {author} {\bibfnamefont {Y.-K.}\ \bibnamefont {Han}}, \ and\
  \bibinfo {author} {\bibfnamefont {Y.~S.}\ \bibnamefont {Huh}},\ }\href
  {\doibase 10.1021/nn305750s} {\bibfield  {journal} {\bibinfo  {journal} {ACS
  Nano}\ }\textbf {\bibinfo {volume} {7}},\ \bibinfo {pages} {2453} (\bibinfo
  {year} {2013})}\BibitemShut {NoStop}%
\bibitem [{\citenamefont {Ghosh}\ \emph {et~al.}(2015)\citenamefont {Ghosh},
  \citenamefont {Mandal},\ and\ \citenamefont {Das}}]{2015GHO}%
  \BibitemOpen
  \bibfield  {author} {\bibinfo {author} {\bibfnamefont {D.}~\bibnamefont
  {Ghosh}}, \bibinfo {author} {\bibfnamefont {M.}~\bibnamefont {Mandal}}, \
  and\ \bibinfo {author} {\bibfnamefont {C.~K.}\ \bibnamefont {Das}},\ }\href
  {\doibase 10.1021/acs.langmuir.5b00649} {\bibfield  {journal} {\bibinfo
  {journal} {Langmuir}\ }\textbf {\bibinfo {volume} {31}},\ \bibinfo {pages}
  {7835} (\bibinfo {year} {2015})}\BibitemShut {NoStop}%
\bibitem [{\citenamefont {Liu}\ \emph {et~al.}(2021)\citenamefont {Liu},
  \citenamefont {Chen}, \citenamefont {Ma}, \citenamefont {bo~Xiong},
  \citenamefont {rong Zeng},\ and\ \citenamefont {Qian}}]{2021LIU}%
  \BibitemOpen
  \bibfield  {author} {\bibinfo {author} {\bibfnamefont {Q.}~\bibnamefont
  {Liu}}, \bibinfo {author} {\bibfnamefont {Y.}~\bibnamefont {Chen}}, \bibinfo
  {author} {\bibfnamefont {J.}~\bibnamefont {Ma}}, \bibinfo {author}
  {\bibfnamefont {X.}~\bibnamefont {bo~Xiong}}, \bibinfo {author}
  {\bibfnamefont {X.}~\bibnamefont {rong Zeng}}, \ and\ \bibinfo {author}
  {\bibfnamefont {H.}~\bibnamefont {Qian}},\ }\href {\doibase
  10.1016/j.surfcoat.2021.127452} {\ \textbf {\bibinfo {volume} {421}},\
  \bibinfo {pages} {127452} (\bibinfo {year} {2021})}\BibitemShut {NoStop}%
\bibitem [{\citenamefont {Bhojane}\ \emph {et~al.}(2019)\citenamefont
  {Bhojane}, \citenamefont {Bail},\ and\ \citenamefont {Shirage}}]{2019BHO}%
  \BibitemOpen
  \bibfield  {author} {\bibinfo {author} {\bibfnamefont {P.}~\bibnamefont
  {Bhojane}}, \bibinfo {author} {\bibfnamefont {A.~L.}\ \bibnamefont {Bail}}, \
  and\ \bibinfo {author} {\bibfnamefont {P.~M.}\ \bibnamefont {Shirage}},\
  }\href {\doibase 10.1107/s2053229618017734} {\bibfield  {journal} {\bibinfo
  {journal} {Acta Crystallographica Section C Structural Chemistry}\ }\textbf
  {\bibinfo {volume} {75}},\ \bibinfo {pages} {61} (\bibinfo {year}
  {2019})}\BibitemShut {NoStop}%
\bibitem [{\citenamefont {Porta}\ \emph {et~al.}(1992)\citenamefont {Porta},
  \citenamefont {Dragone}, \citenamefont {Fierro}, \citenamefont {Inversi},
  \citenamefont {Jacono},\ and\ \citenamefont {Moretti}}]{1992POR}%
  \BibitemOpen
  \bibfield  {author} {\bibinfo {author} {\bibfnamefont {P.}~\bibnamefont
  {Porta}}, \bibinfo {author} {\bibfnamefont {R.}~\bibnamefont {Dragone}},
  \bibinfo {author} {\bibfnamefont {G.}~\bibnamefont {Fierro}}, \bibinfo
  {author} {\bibfnamefont {M.}~\bibnamefont {Inversi}}, \bibinfo {author}
  {\bibfnamefont {M.~L.}\ \bibnamefont {Jacono}}, \ and\ \bibinfo {author}
  {\bibfnamefont {G.}~\bibnamefont {Moretti}},\ }\href {\doibase
  10.1039/ft9928800311} {\bibfield  {journal} {\bibinfo  {journal} {J. Chem.
  Soc., Faraday Trans.}\ }\textbf {\bibinfo {volume} {88}},\ \bibinfo {pages}
  {311} (\bibinfo {year} {1992})}\BibitemShut {NoStop}%
\bibitem [{\citenamefont {Simon}\ and\ \citenamefont
  {Gogotsi}(2008)}]{2008SIM}%
  \BibitemOpen
  \bibfield  {author} {\bibinfo {author} {\bibfnamefont {P.}~\bibnamefont
  {Simon}}\ and\ \bibinfo {author} {\bibfnamefont {Y.}~\bibnamefont
  {Gogotsi}},\ }\href {\doibase 10.1038/nmat2297} {\ \textbf {\bibinfo {volume}
  {7}},\ \bibinfo {pages} {845} (\bibinfo {year} {2008})}\BibitemShut {NoStop}%
\bibitem [{\citenamefont {Kim}\ \emph {et~al.}(2018)\citenamefont {Kim},
  \citenamefont {Kim}, \citenamefont {Kim}, \citenamefont {Sadan},
  \citenamefont {Yeo}, \citenamefont {Cho}, \citenamefont {Ahn}, \citenamefont
  {Ahn},\ and\ \citenamefont {Ahn}}]{2018KIM}%
  \BibitemOpen
  \bibfield  {author} {\bibinfo {author} {\bibfnamefont {C.}~\bibnamefont
  {Kim}}, \bibinfo {author} {\bibfnamefont {I.}~\bibnamefont {Kim}}, \bibinfo
  {author} {\bibfnamefont {H.}~\bibnamefont {Kim}}, \bibinfo {author}
  {\bibfnamefont {M.~K.}\ \bibnamefont {Sadan}}, \bibinfo {author}
  {\bibfnamefont {H.}~\bibnamefont {Yeo}}, \bibinfo {author} {\bibfnamefont
  {G.}~\bibnamefont {Cho}}, \bibinfo {author} {\bibfnamefont {J.}~\bibnamefont
  {Ahn}}, \bibinfo {author} {\bibfnamefont {J.}~\bibnamefont {Ahn}}, \ and\
  \bibinfo {author} {\bibfnamefont {H.}~\bibnamefont {Ahn}},\ }\href {\doibase
  10.1039/c8ta09544b} {\ \textbf {\bibinfo {volume} {6}},\ \bibinfo {pages}
  {22809} (\bibinfo {year} {2018})}\BibitemShut {NoStop}%
\bibitem [{\citenamefont {Toma}\ \emph {et~al.}(2020)\citenamefont {Toma},
  \citenamefont {Maezono},\ and\ \citenamefont {Hongo}}]{2020TOM}%
  \BibitemOpen
  \bibfield  {author} {\bibinfo {author} {\bibfnamefont {T.}~\bibnamefont
  {Toma}}, \bibinfo {author} {\bibfnamefont {R.}~\bibnamefont {Maezono}}, \
  and\ \bibinfo {author} {\bibfnamefont {K.}~\bibnamefont {Hongo}},\ }\href
  {\doibase 10.1021/acsaem.0c00602} {\ \textbf {\bibinfo {volume} {3}},\
  \bibinfo {pages} {4078} (\bibinfo {year} {2020})}\BibitemShut {NoStop}%
\bibitem [{\citenamefont {Xu}\ and\ \citenamefont {Zeng}(2003)}]{2003XU}%
  \BibitemOpen
  \bibfield  {author} {\bibinfo {author} {\bibfnamefont {R.}~\bibnamefont
  {Xu}}\ and\ \bibinfo {author} {\bibfnamefont {H.~C.}\ \bibnamefont {Zeng}},\
  }\href {\doibase 10.1021/jp035751c} {\bibfield  {journal} {\bibinfo
  {journal} {The Journal of Physical Chemistry B}\ }\textbf {\bibinfo {volume}
  {107}},\ \bibinfo {pages} {12643} (\bibinfo {year} {2003})}\BibitemShut
  {NoStop}%
\bibitem [{\citenamefont {Momma}\ and\ \citenamefont {Izumi}(2011)}]{2011MOM}%
  \BibitemOpen
  \bibfield  {author} {\bibinfo {author} {\bibfnamefont {K.}~\bibnamefont
  {Momma}}\ and\ \bibinfo {author} {\bibfnamefont {F.}~\bibnamefont {Izumi}},\
  }\href {\doibase 10.1107/s0021889811038970} {\bibfield  {journal} {\bibinfo
  {journal} {J Appl Cryst}\ }\textbf {\bibinfo {volume} {44}},\ \bibinfo
  {pages} {1272} (\bibinfo {year} {2011})}\BibitemShut {NoStop}%
\bibitem [{\citenamefont {{CRC Handbook}}(2007)}]{2007LID}%
  \BibitemOpen
  \bibfield  {author} {\bibinfo {author} {\bibnamefont {{CRC Handbook}}},\
  }\href
  {http://www.amazon.com/CRC-Handbook-Chemistry-Physics-88th/dp/0849304881/ref=sr_1_5?ie=UTF8&qid=1302802093&sr=8-5}
  {\emph {\bibinfo {title} {CRC Handbook of Chemistry and Physics, 88th
  Edition}}},\ \bibinfo {edition} {88th}\ ed.\ (\bibinfo  {publisher} {CRC
  Press},\ \bibinfo {year} {2007})\BibitemShut {NoStop}%
\bibitem [{\citenamefont {Kresse}\ and\ \citenamefont
  {Furthm\"uller}(1996)}]{1996KRE}%
  \BibitemOpen
  \bibfield  {author} {\bibinfo {author} {\bibfnamefont {G.}~\bibnamefont
  {Kresse}}\ and\ \bibinfo {author} {\bibfnamefont {J.}~\bibnamefont
  {Furthm\"uller}},\ }\href {\doibase 10.1103/PhysRevB.54.11169} {\bibfield
  {journal} {\bibinfo  {journal} {Phys. Rev. B}\ }\textbf {\bibinfo {volume}
  {54}},\ \bibinfo {pages} {11169} (\bibinfo {year} {1996})}\BibitemShut
  {NoStop}%
\bibitem [{\citenamefont {Kresse}\ and\ \citenamefont
  {Joubert}(1999)}]{1999KRE}%
  \BibitemOpen
  \bibfield  {author} {\bibinfo {author} {\bibfnamefont {G.}~\bibnamefont
  {Kresse}}\ and\ \bibinfo {author} {\bibfnamefont {D.}~\bibnamefont
  {Joubert}},\ }\href {\doibase 10.1103/PhysRevB.59.1758} {\bibfield  {journal}
  {\bibinfo  {journal} {Phys. Rev. B}\ }\textbf {\bibinfo {volume} {59}},\
  \bibinfo {pages} {1758} (\bibinfo {year} {1999})}\BibitemShut {NoStop}%
\bibitem [{\citenamefont {Bl\"ochl}(1994)}]{1994BLO}%
  \BibitemOpen
  \bibfield  {author} {\bibinfo {author} {\bibfnamefont {P.~E.}\ \bibnamefont
  {Bl\"ochl}},\ }\href {\doibase 10.1103/PhysRevB.50.17953} {\bibfield
  {journal} {\bibinfo  {journal} {Phys. Rev. B}\ }\textbf {\bibinfo {volume}
  {50}},\ \bibinfo {pages} {17953} (\bibinfo {year} {1994})}\BibitemShut
  {NoStop}%
\bibitem [{\citenamefont {Perdew}\ \emph {et~al.}(1996)\citenamefont {Perdew},
  \citenamefont {Burke},\ and\ \citenamefont {Ernzerhof}}]{1996PER}%
  \BibitemOpen
  \bibfield  {author} {\bibinfo {author} {\bibfnamefont {J.~P.}\ \bibnamefont
  {Perdew}}, \bibinfo {author} {\bibfnamefont {K.}~\bibnamefont {Burke}}, \
  and\ \bibinfo {author} {\bibfnamefont {M.}~\bibnamefont {Ernzerhof}},\ }\href
  {\doibase 10.1103/physrevlett.77.3865} {\bibfield  {journal} {\bibinfo
  {journal} {Physical Review Letters}\ }\textbf {\bibinfo {volume} {77}},\
  \bibinfo {pages} {3865} (\bibinfo {year} {1996})}\BibitemShut {NoStop}%
\bibitem [{\citenamefont {Wdowik}\ and\ \citenamefont
  {Parlinski}(2007)}]{2007WDO}%
  \BibitemOpen
  \bibfield  {author} {\bibinfo {author} {\bibfnamefont {U.~D.}\ \bibnamefont
  {Wdowik}}\ and\ \bibinfo {author} {\bibfnamefont {K.}~\bibnamefont
  {Parlinski}},\ }\href {\doibase 10.1103/PhysRevB.75.104306} {\bibfield
  {journal} {\bibinfo  {journal} {Phys. Rev. B}\ }\textbf {\bibinfo {volume}
  {75}},\ \bibinfo {pages} {104306} (\bibinfo {year} {2007})}\BibitemShut
  {NoStop}%
\bibitem [{\citenamefont {Grimme}\ \emph {et~al.}(2010)\citenamefont {Grimme},
  \citenamefont {Antony}, \citenamefont {Ehrlich},\ and\ \citenamefont
  {Krieg}}]{2010GRI}%
  \BibitemOpen
  \bibfield  {author} {\bibinfo {author} {\bibfnamefont {S.}~\bibnamefont
  {Grimme}}, \bibinfo {author} {\bibfnamefont {J.}~\bibnamefont {Antony}},
  \bibinfo {author} {\bibfnamefont {S.}~\bibnamefont {Ehrlich}}, \ and\
  \bibinfo {author} {\bibfnamefont {H.}~\bibnamefont {Krieg}},\ }\href
  {\doibase 10.1063/1.3382344} {\bibfield  {journal} {\bibinfo  {journal} {The
  Journal of Chemical Physics}\ }\textbf {\bibinfo {volume} {132}},\ \bibinfo
  {pages} {154104} (\bibinfo {year} {2010})}\BibitemShut {NoStop}%
\bibitem [{\citenamefont {Johnson}\ and\ \citenamefont
  {Becke}(2006)}]{2006JOH}%
  \BibitemOpen
  \bibfield  {author} {\bibinfo {author} {\bibfnamefont {E.~R.}\ \bibnamefont
  {Johnson}}\ and\ \bibinfo {author} {\bibfnamefont {A.~D.}\ \bibnamefont
  {Becke}},\ }\href {\doibase 10.1063/1.2190220} {\bibfield  {journal}
  {\bibinfo  {journal} {The Journal of Chemical Physics}\ }\textbf {\bibinfo
  {volume} {124}},\ \bibinfo {pages} {174104} (\bibinfo {year}
  {2006})}\BibitemShut {NoStop}%
\bibitem [{\citenamefont {Monkhorst}\ and\ \citenamefont
  {Pack}(1976)}]{1976MON}%
  \BibitemOpen
  \bibfield  {author} {\bibinfo {author} {\bibfnamefont {H.~J.}\ \bibnamefont
  {Monkhorst}}\ and\ \bibinfo {author} {\bibfnamefont {J.~D.}\ \bibnamefont
  {Pack}},\ }\href {\doibase 10.1103/PhysRevB.13.5188} {\bibfield  {journal}
  {\bibinfo  {journal} {Phys. Rev. B}\ }\textbf {\bibinfo {volume} {13}},\
  \bibinfo {pages} {5188} (\bibinfo {year} {1976})}\BibitemShut {NoStop}%
\bibitem [{\citenamefont {Gaunand}\ and\ \citenamefont {Lim}(2002)}]{2002GAU}%
  \BibitemOpen
  \bibfield  {author} {\bibinfo {author} {\bibfnamefont {A.}~\bibnamefont
  {Gaunand}}\ and\ \bibinfo {author} {\bibfnamefont {W.}~\bibnamefont {Lim}},\
  }\href {\doibase 10.1016/s0032-5910(02)00276-0} {\bibfield  {journal}
  {\bibinfo  {journal} {Powder Technology}\ }\textbf {\bibinfo {volume}
  {128}},\ \bibinfo {pages} {332} (\bibinfo {year} {2002})}\BibitemShut
  {NoStop}%
\end{thebibliography}%

%%%%%%%%%% Merge with supplemental materials %%%%%%%%%%
\newpage
\begin{center}
\textbf{\large Supplemental Materials:  First-Principles-Based Insight into Electrochemical Reactivity in a Cobalt-Carbonate-Hydrate Pseudocapacitor}
\end{center}
%%%%%%%%%% Merge with supplemental materials %%%%%%%%%%
%%%%%%%%%% Prefix a "S" to all equations, figures, tables and reset the counter %%%%%%%%%%
\setcounter{equation}{0}
\setcounter{figure}{0}
\setcounter{table}{0}
\setcounter{page}{1}
\renewcommand{\thefigure}{S-\arabic{figure}}
\makeatletter

%%%%%%%%%%%%%%%%%%%%%%%%%%%%%%%%%%%%%%%%%%%%%%%%%%%%%
\section{Hydrogen bonding between specific planes.}
%%%%%%%%%%%%%%%%%%%%%%%%%%%%%%%%%%%%%%%%%%%%%%%%%%%%%
NA-CCH and CA-CCH are known as Cobalt Carbonate Hydroxide (CCH) electrode materials
which have different crystal growth directions
(NA-CCH : one-dimensional, CA-CCH : two-dimensional).
Both materaials have very different capacitor properties due to its molphology difference.
We found that the difference in crystal growth direction
can be explained by the hydrogen bonding direction.
For NA-CCH, more number of the three crystal planes ($2\theta=33.515$) in Figure S 2 are formed 
than that of the three crystal planes ($2\theta=9.885$) in Figure S 1.
On the other hand, CA-CCH shows opposite trend.
Therefore, the hydrogen bonding of NA-CCH would be dominantly formed to $c$-axis direction
than $ab$ inner plane direction.
Conversely, hydrogen bonding in the $ab$ inner plane direction
would be dominant for CA-CCH.

\vspace{2mm}
Furthermore, this difference of hydrogen bonding
direction would be explained by
the difference of occupancy ratio of structural water site.
As shown the $c$-axis directional hydrogen bonding in Figure S2,
structural water site (orange-colored ellipses) forms two hydrogen bonding
between ($\bar{1}\bar{1}\bar{1}$) plane.
NA-CCH has higher occupancy ratio of the structural water site
than CA-CCH.
Therefore, hydrogen bonding site provided by structural water site
would promote one directional crystal growth of NA-CCH.

%--------------------------------
\begin{figure*}[h]
  \centering
  \includegraphics[width=0.8
    \hsize]{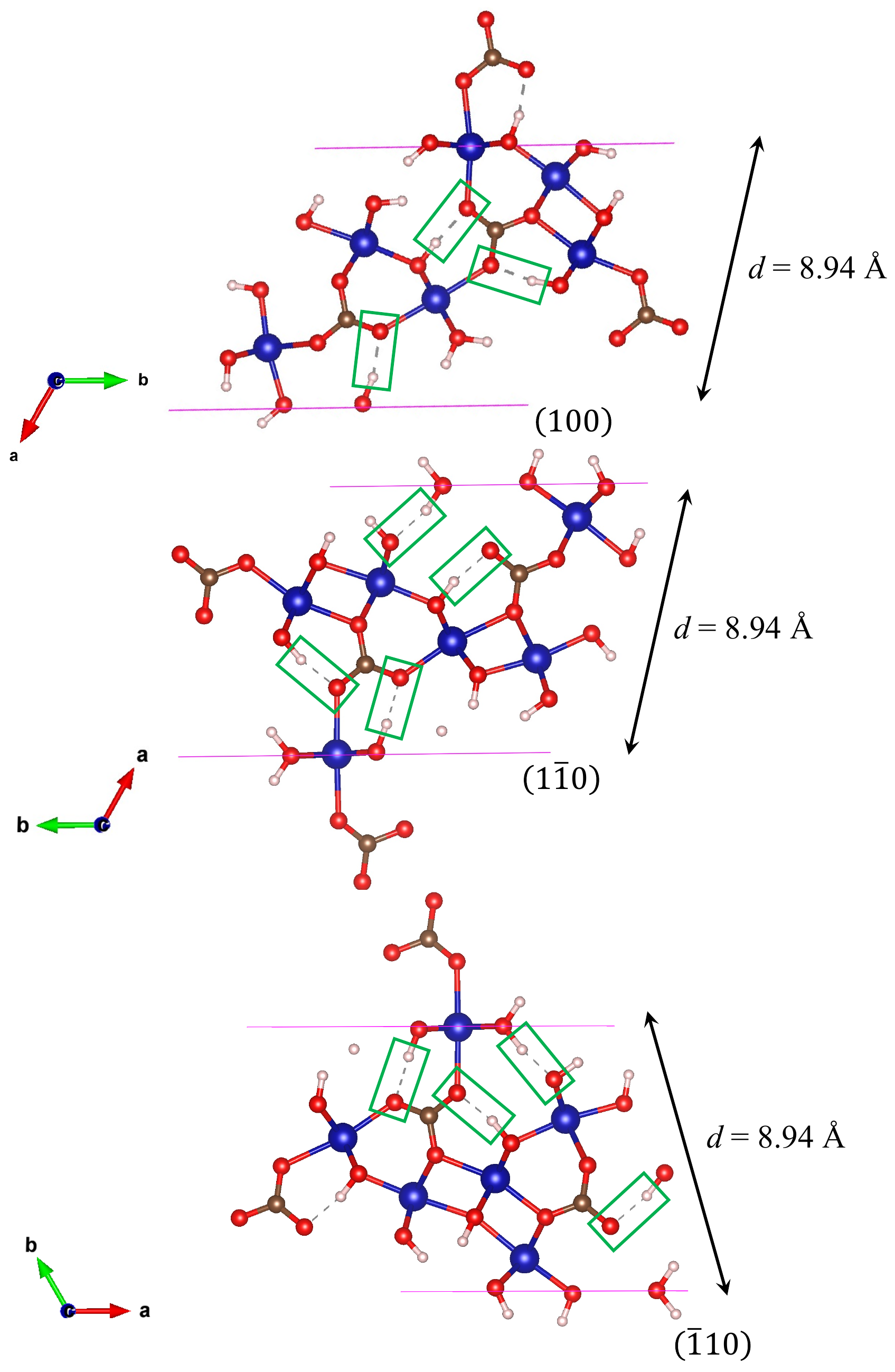}
  \caption{
    \label{fig.ab_bonding_plane}
    Hydrogen bonding in the geometry optimized CCH structure.
    Hydrogen bonding sites are enclosed by green rectangle.
    Hydrogen bondings which formed between each plane (100), ($\bar{1}$10), (1$\bar{1}$0)
    are nearly vertical to $\textit{ab}$ plane.
    These 2 lattice planes correspond to the XRD peak at 2$\theta$=9.885.
  }
\end{figure*}
%--------------------------------

%%--------------------------------
%\begin{figure*}[h]
%  \centering
%  \includegraphics[width=0.8
%    \hsize]{./si_fig/ab_bonding2_plane.pdf}
%  \caption{
%    \label{fig.ab_bonding2_plane}
%    {\color{red}
%    Hydrogen bonding in the geometry optimized in CCH strucutre.
%    Hydrogen bonding sites are enclosed by green rectangle.
%    Hydrogen bondings which formed between each plane ($1\bar{2}0$), ($\bar{2}$10)
%    are nearly parallel to $\textit{ab}$ plane.
%    These 2 lattice planes correspond to the XRD peak at 2$\theta$=17.165.
%    }
%  }
%\end{figure*}
%%--------------------------------

%--------------------------------
\begin{figure*}[h]
  \centering
  \includegraphics[width=0.99\hsize]{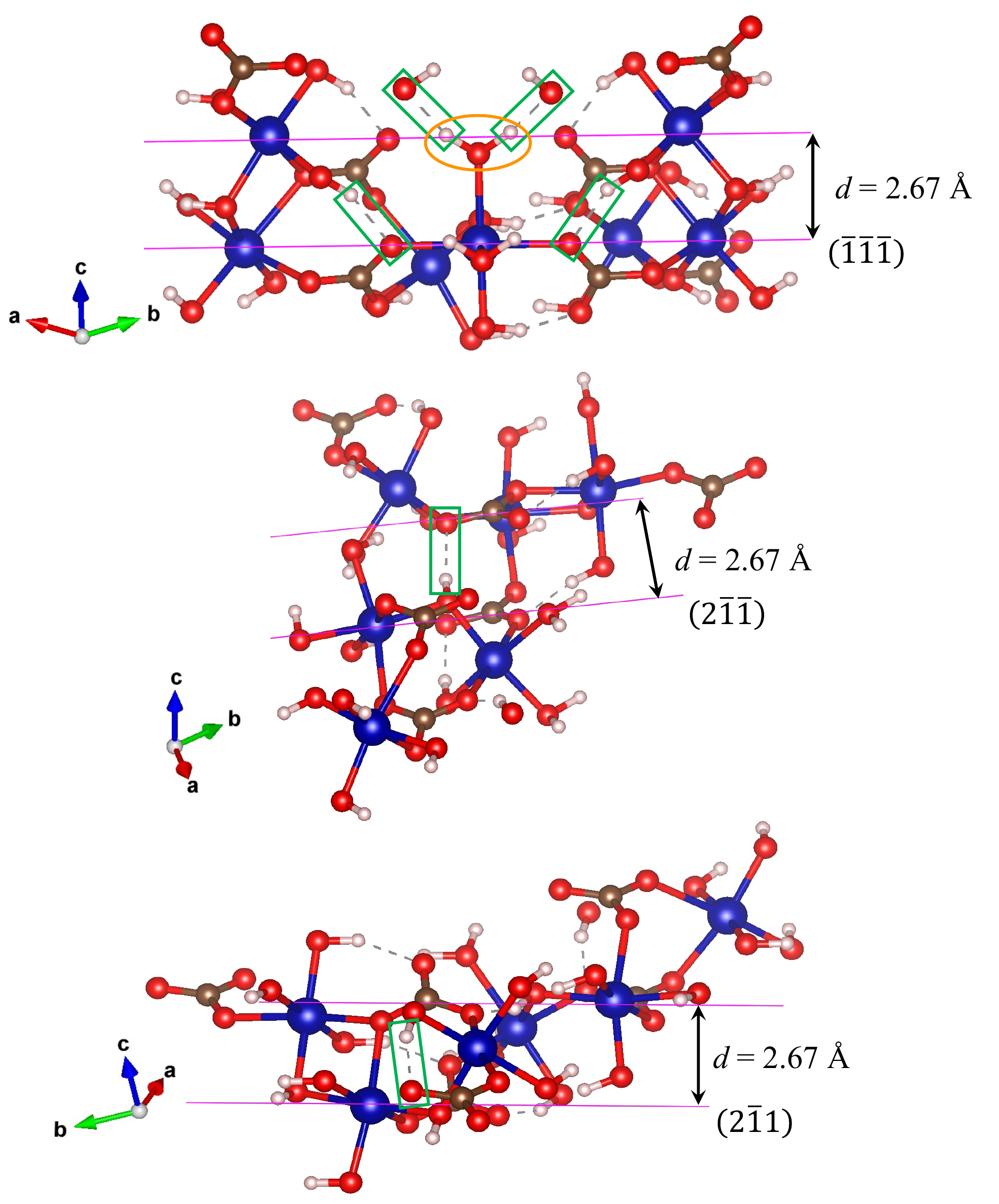}
  \caption{
    \label{fig.c_bonding_plane}
    Hydrogen bonding in the geometry optimized CCH structure.
    Hydrogen bonding sites are enclosed by green rectangle.
    The directions of hydrogen bondings formed between each plane
    ($\bar{1}$$\bar{1}$$\bar{1}$), (2$\bar{1}$$\bar{1}$), and (2$\bar{1}$1)
    are close to $\textit{c}$ axis direction.
    These 3 lattice planes correspond to the XRD peak at 2$\theta$=33.515.
    Two hydrogen bonding provided by structural water (orange-colored ellipses)
    are formed between ($\bar{1}$$\bar{1}$$\bar{1}$) planes.
  }
\end{figure*}
%--------------------------------

\end{document}